\documentclass[aps,twocolumn,nofootinbib]{revtex4}

\usepackage{epic}
\usepackage{graphicx}

\usepackage[T1]{fontenc}
\usepackage{amssymb,amsmath,amsfonts}
\usepackage{stackrel}
\usepackage{float}

\usepackage{color}
\usepackage{ulem}

\newcommand{\Ayan}[1]{{\color[rgb]{0,0,.0}#1}}

\newcommand{\lata}[1]{{\color[rgb]{0,0,0}#1}}

\newcommand{\be}{\begin{equation}}
\newcommand{\ee}{\end{equation}}
\newcommand{\bse}{\begin{subequations}}
\newcommand{\ese}{\end{subequations}}
\newcommand{\ba}{\begin{eqnarray}}
\newcommand{\ea}{\end{eqnarray}}
\newcommand{\bea}{\begin{eqnarray}}
\newcommand{\eea}{\end{eqnarray}}
\usepackage{subfig}
\newcommand{\pb}{\Phi_{\rm b}}

\begin{document}

\title{Time-dependent $NAdS_2$ holography with applications}
\author{Lata Kh Joshi}
\email{latakj@theory.tifr.res.in}
\affiliation{ Department of Theoretical Physics, Tata Institute of Fundamental Research, Homi Bhabha Road, Bombay 400 005, India}
\author{Ayan Mukhopadhyay} 
\email{ayan@iitm.ac.in}
\affiliation{Department of Physics, Indian Institute of Technology Madras, Chennai 600036, India}
\author{Alexander Soloviev}
\email{alexander.soloviev@tuwien.ac.at}
\affiliation{Institut f\"{u}r Theoretische Physik, Technische Universit\"{a}t Wien, Wiedner Hauptstr.~8-10, A-1040 Vienna, Austria}

\begin{abstract}
We develop a method for obtaining exact time-dependent solutions in Jackiw-Teitelboim gravity coupled to non-conformal matter and study consequences for $NAdS_2$ holography. {We study holographic quenches in which we find that the black hole mass increases.} A semi-holographic model composed of an infrared $NAdS_2$ holographic sector representing the mutual strong interactions of trapped impurities confined at a spatial point is proposed. The holographic sector couples to the position of a displaced impurity acting as a self-consistent boundary source. This effective $0+1-$dimensional description has a total conserved energy. {Irrespective of the initial velocity of the particle, the black hole mass initially increases, but after the horizon runs away to infinity in the physical patch, the mass vanishes in the long run. The total energy is completely transferred to the kinetic energy or the self-consistent confining potential energy of the impurity.} For initial velocities below a critical value determined by the mutual coupling, \Ayan{ {the black hole mass changes sign in finite time. Above this critical velocity, the initial condition of the particle can be retrieved from the  $SL(2,R)$ invariant exponent that governs the exponential growth of the bulk gravitational $SL(2,R)$ charges at late time.  }}
\end{abstract}

\maketitle

\section{Introduction}
Nearly-$AdS_2$ holography \cite{Almheiri:2014cka,Jensen:2016pah,Maldacena:2016upp,Engelsoy:2016xyb} provides a rich playground for exploring many fundamental questions. It is a template for understanding the inner working of the holographic correspondence given that some possible dual systems, such as the infrared regime of the SYK model, can also be exactly solved in the large $N$ limit \cite{Sachdev:1992fk,Kitaev:2017awl} (see also \cite{Gross:2016kjj,Witten:2016iux,Klebanov:2016xxf}). Other significant applications are the deeper understanding of real-time holography that can shed new light on quantum many-body systems (especially those which are maximally chaotic), and also new insights on the black hole information loss paradox via a solvable toy model of real-time black hole evaporation. For such applications, the setup of nearly-$AdS_2$ holography has to include additional bulk fields which provide propagating modes in the $2D$ gravity theory. In this work, we explore such setups in the classical regime in real time by generating exact time-dependent solutions.

Since nearly-$AdS_2$ holographic systems have an intrinsic cut-off scale, it is well motivated to study a semi-holographic setup where the holographic degrees of freedom are coupled to a dynamical source at the boundary of $AdS_2$ depicting UV dynamics. The mutual coupling must be such that the total energy is conserved. Semi-holographic constructions of this type have been explored in the context of non-Fermi liquids \cite{Faulkner:2010tq,Mukhopadhyay:2013dqa,Doucot:2017bdm} and also the quark-gluon plasma \cite{Iancu:2014ava,Mukhopadhyay:2015smb,Kurkela:2018dku,Ecker:2018ucc}. 

\Ayan{The semi-holographic approach is suited for scenarios where a holographic description can be valid only for some strongly interacting infrared degrees of freedom of the system whilst some degrees of freedom are perturbative. It can be relevant for understanding the formation of quark-gluon plasma from perturbative processes \cite{Iancu:2014ava}. A fundamental derivation of semi-holography in the context of QCD has been discussed in \cite{Banerjee:2017ozx}. From the point of view of phenomenological applications, it also gives us a flexible way to apply holography to laboratory setups where the ultraviolet complete description is not relevant \cite{Faulkner:2010tq,Faulkner:2010jy}. It is from this perspective that we apply semi-holographic approach for a nearly-$AdS_2$ holographic system coupled to a dynamical source at the boundary. We apply this to study confined strongly interacting impurities. }

 In our model, the $NAdS_2$ holographic sector depicts the dual infrared dynamics of many-body interactions localized at the origin where the impurities are confined. The motion in space of an impurity can be thought of as a deformation of this \Ayan{ $0+1-$dimensional} $NAdS_2$ holographic theory with the time-dependent position of the impurity representing a self-consistent external source \Ayan{of an irrelevant operator with a dynamically generated expectation value}. The displaced impurity in turn follows Newtonian law of motion under the influence of the force generated by its coupling to the bulk field -- \Ayan{the dual irrelevant holographic operator now generates the tension of the confining force. Since the $NAdS_2$ holographic sector is an infrared conformal theory, it should be deformed only via an irrelevant operator. The semi-holographic setup then models the dynamics at intermediate energy scales phenomenologically such that the total energy of the system is always conserved. We study the exact time-dependent solutions of the full system in this model.}
 
The gravitational description for nearly-$AdS_2$ holography is the two-dimensional Jackiw-Teitelboim (JT) gravity with non-conformal matter \cite{Teitelboim:1983ux,Jackiw:1984je,Brown:1988am}. The peculiarity of the JT model is that the metric is always locally $AdS_2$. This is ensured by the presence of a non-propagating dilaton field, which does not couple to matter. Its equation of motion enforces the Ricci scalar to be a constant, acting like a Lagrange multiplier. Nevertheless, the dilaton's boundary condition even in the absence of matter generates non-trivial states in the dual theory, which can be characterized by time-reparametrizations just like in the SYK model.  This JT gravity coupled to matter cannot be lifted to a higher dimensional setup, because if it were possible, then the dilaton which maps to the size of the extra compactified space should have coupled to matter (see \cite{Mertens:2018fds,Gaikwad:2018dfc,Nayak:2018qej,Larsen:2018iou} and also \cite{Das:2017pif} for instance).

The presence of a second law of thermodynamics in JT gravity coupled to matter turns out to be a subtle issue \cite{Frolov:1992xx,Duchting:2000qk}. The formation of a horizon alone does not guarantee a second law as in the case of higher dimensional setups. Under the right circumstances, the value of the dilaton on the horizon grows monotonically {and essentially coincides with the thermal entropy when the system thermalizes. However, we will find that in the semi-holographic setup the runaway behavior of the horizon and entanglement of the particle with the coarse-grained macroscopic variables of the holographic system leads to a novel scenario where the black hole mass can vanish without any contradiction with the second law.}

We find remarkably that although in pure holographic setups the mass of a pre-existing black hole increases when subjected to a quench at the boundary, in our semi-holographic model the pre-existing black hole is always completely depleted of its mass at long time. This behavior is the reverse of what we find in higher dimensional semi-holographic setups in the presence of scalar mutual couplings.\footnote{A similar phenomenon of disappearance of horizon in the bulk in nearly $AdS_2$ setups has been found in \cite{Kourkoulou:2017zaj}. The explanation proposed for this result also works naturally in our case.} 

The plan of the paper is as follows. In Section 2, we discuss JT gravity coupled to matter and its holographic interpretation. To be self-contained, we present results previously obtained by other authors along with some new ones. In Section 3, we present our method for obtaining exact time-dependent solutions in JT gravity and study quenches in the pure holographic setup. In Section 4, we present the semi-holographic model for impurities and study its solutions. In Section 5, we conclude with a summary of results, and discuss 
some open questions.

\section{The setup}
\subsection{Bulk equations of motion}
The Jackiw-Teitelboim model \cite{Teitelboim:1983ux,Jackiw:1984je,Brown:1988am} provides the simplest example of a two-dimensional pure gravity theory in which non-vacuum states with finite energy exist. To produce time-dependent solutions though, we need to have either time-dependent boundary conditions or couple it to self-interacting matter. 

The general version of the action which is suitable for taking the large $N$ type limit in the dual theory is
\begin{align}
S &= \frac{1}{16\pi G}\left[ \int {\rm d}^2 x \sqrt{-g}\Phi\left(R + \frac{2}{l^2}\right) + S_{\rm matter}[g, \chi]\right]\nonumber\\
&+ \frac{1}{8\pi G} \int {\rm d}u\,\sqrt{-h}\, \pb K .
\end{align}
Note that $u$ appearing in the Gibbons-Hawking-York counterterm above is to be identified with the \textit{boundary time} i.e. the time of the boundary observer. Also $\Phi_b$ is simply the value of $\Phi$ at the boundary. The key feature of this theory is that the dilaton field $\Phi$ does \textit{not} couple to matter. This implies that the bulk metric remains always pure $AdS_2$ \textit{locally}. Indeed by varying the action with respect to $\Phi$, we simply obtain
\be 
R +\frac{2}{l^2} =0.
\ee
Varying the action with respect to the bulk metric yields 
\be\label{PhiCons}
T_{\mu\nu}^\Phi  + T_{\mu\nu} = 0,
\ee
where
\begin{align}
T_{\mu\nu}^\Phi &\equiv \nabla_\mu \nabla_\nu \Phi - g_{\mu\nu} \nabla^2 \Phi +\frac{1}{l^2} g_{\mu\nu} \Phi, \nonumber\\
T_{\mu\nu} &= - \frac{2}{\sqrt{-g}} \frac{\delta S_{\rm matter}}{\delta g^{\mu\nu}}.
\end{align}
Note that the Bianchi identity is satisfied when $R = - 2/l^2$. Therefore, the equation of motion \eqref{PhiCons} is indeed consistent in a locally $AdS_2$ background spacetime. We set $l=1$ by appropriate choice of units for the sake of our convenience.

In what follows, we will \textit{not} assume that the matter sector is conformal and thus generalize the results in \cite{Almheiri:2014cka,Engelsoy:2016xyb}. Without assuming specific details of the matter sector, we can readily proceed by only implementing the local conservation of energy and momentum, i.e.
\be\label{consv}
\nabla_\mu T^{\mu\nu} = 0.
\ee
Since the background spacetime remains locally $AdS_2$, we can always adopt the  Fefferman-Graham coordinates
\be\label{PPform}
{\rm d}s^2 = \frac{1}{z^2}\left(-{\rm d}t^2 + {\rm d}z^2\right).
\ee
in which the conservation equations assume the explicit form
\be
\partial_z T_{zt} = \partial_t{T}_{tt} , \quad \partial_z (zT_{zz}) = T_{tt} + z\, \partial_t{T}_{zt}.
\ee
The general form of $T_{\mu\nu}$ should then be
\begin{align}\label{Tztcons}
T_{zt}(z,t) &= F_\epsilon(t) + \int_\epsilon^z{\rm d}z_1\,\, \partial_t{T}_{tt}(z_1, t) , \\\nonumber
T_{zz}(z,t) &= \frac{G_\epsilon(t)}{z}  + \frac{z}{2}\partial_t{F}_\epsilon(t)+\frac{1}{z}\int_\epsilon^z{\rm d}z_1\,\,  {T}_{tt}(z_1, t) \nonumber\\&+ \frac{z}{2} \int_\epsilon^z{\rm d}z_1\,\, \partial_t^2{T}_{tt}(z_1,t)\nonumber\\
&-\frac{1}{2z}\int_\epsilon^z{\rm d}z_1\,\, \partial_t^2{T}_{tt}(z_1,t) z_1^2,\label{Tzzcons}
\end{align}
with
\begin{eqnarray}\label{F}
F_\epsilon(t) &=& T_{zt}(\epsilon, t),\\\label{G}
G_\epsilon(t) &=& \epsilon T_{zz}(\epsilon, t)- \frac{\epsilon^2}{2}\partial_t{T}_{zt}(\epsilon, t)
\end{eqnarray}
and $\epsilon$ denoting an arbitrary radial cut-off which for the sake of convenience should be chosen close to the \textit{boundary} $z= 0$. The boundary conditions for the bulk matter fields determine $F_\epsilon(t)$ and $G_\epsilon(t)$.  Then \eqref{Tztcons} and \eqref{Tzzcons} determine $T_{zt}$ and $T_{zz}$  in terms of $T_{tt}$.

The various components of \eqref{PhiCons} turn out to be
\begin{eqnarray}\label{phitteom}
\partial_z^2\Phi+ \frac{\partial_z\Phi}{z} - \frac{\Phi}{z^2} =- T_{tt}, \\\label{constraint1}
 \partial_z\partial_t{\Phi} +\frac{\partial_t{\Phi}}{z}= -T_{zt},\\\label{constraint2}
\partial_t^2{\Phi} + \frac{\partial_z\Phi}{z}+ \frac{\Phi}{z^2} = -T_{zz}
\end{eqnarray}
in the Fefferman-Graham coordinates. The first equation above involving only radial derivatives determines the radial profile of $\Phi$. The remaining equations reduce simply to constraints after we utilize the results of matter energy and momentum conservation given by \eqref{Tztcons} and \eqref{Tzzcons}. These constraints are therefore only time-dependent equations determining data at the cut-off $z = \epsilon$.

The most general solution of \eqref{phitteom} is
\begin{align}\label{phisol1}
\Phi(z,t) &= \frac{\alpha_\epsilon(t)}{z} + \beta_\epsilon(t) z - \frac{z}{2} \int_\epsilon^z{\rm d}z_1\,\,   {T}_{tt}(z_1, t) \nonumber\\
&+\frac{1}{2z} \int_\epsilon^z{\rm d}z_1\,\,  {T}_{tt}(z_1, t) z_1^2.
\end{align}
Substituting the above in \eqref{constraint1} and \eqref{constraint2}, and also utilizing \eqref{Tztcons} and \eqref{Tzzcons} we obtain
\begin{align}\label{beta-1}
2 \partial_t{\beta}_\epsilon(t) + F_\epsilon (t) = 0, \\\label{alpha-1}
\partial_t^2{\alpha}_\epsilon(t) + 2 \beta_\epsilon(t) + G_\epsilon(t) = 0.
\end{align}
As claimed, these determine the two time-dependent functions in  \eqref{phisol1} and thus the data on the cut-off. Utilizing \eqref{F} and \eqref{G} we obtain the following useful form of these constraints:
\begin{align}\label{beta-2}
\partial_t\beta_\epsilon(t)  &=- \frac{1}{2} T_{zt}(\epsilon, t), \\\label{alpha-2}
\partial_t^3\alpha_\epsilon(t)  &=  T_{zt}(\epsilon, t) + \frac{\epsilon^2}{2}\partial_t^2{T}_{zt}(\epsilon, t)-\epsilon\, \partial_t T_{zz}(\epsilon, t).
\end{align}
Note that \eqref{beta-1} and \eqref{alpha-1} are equivalent to the above only if we choose appropriate integration constants in $\alpha_\epsilon(t)$. This issue is readily addressed if we use the following integral forms for $\alpha_\epsilon(t)$ and $\beta_\epsilon(t)$:
\begin{align}\label{beta-f}
\beta_\epsilon(t)  &=- C_\epsilon- \frac{1}{2}\int_{-\infty}^t {\rm d}t_1 T_{zt}(\epsilon, t_1), \\\label{alpha-f}
\alpha_\epsilon(t)  &= A_\epsilon + B_\epsilon t + C_\epsilon t^2 \nonumber\\
&+\int_{-\infty}^t {\rm d}t_1\int_{-\infty}^{t_1} {\rm d}t_2\int_{-\infty}^{t_2} {\rm d}t_3 \Big[T_{zt}(\epsilon, t_3) 
\nonumber\\&-\epsilon\, \partial_t T_{zz}(\epsilon, t_3)+ \frac{\epsilon^2}{2}\partial_t^2{T}_{zt}(\epsilon, t_3)\Big].
\end{align}
Above $A_\epsilon$, $B_\epsilon$ and $C_\epsilon$ are arbitrary constants. These expressions together with \eqref{phisol1} thus completely specify $\Phi$ in the presence of bulk matter.

\subsection{Holographic interpretation}
The holographic dictionary for the Jackiw-Teitelboim model has been established in \cite{Almheiri:2014cka,Jensen:2016pah,Maldacena:2016upp,Engelsoy:2016xyb} and a thorough treatment of holographic renormalization can be found in \cite{Cvetic:2016eiv,Gonzalez:2018enk} (see also \cite{Grumiller:2007ju}).  In the general situation, it is required to impose an appropriate self-consistent cut-off so that the dual quantum theory lives on an appropriate trajectory $z  = \epsilon f(t)$ which should be determined from the equations of motion themselves. The trajectory of the cut-off does not always coincide with the boundary of the $AdS_2$ spacetime which is at $z = 0$ but is typically \textit{near} the latter if $\epsilon$ is sufficiently small. The dimensionful parameter $\epsilon$ is related to the UV cut-off of the dual quantum theory and is thus an \textit{external} parameter. It turns out that we can take the limit $\epsilon \rightarrow 0$ when the matter sector satisfies certain conditions. In this happy situation, the dual quantum theory is UV complete.\footnote{This statement is true strictly in the large $N$ limit only where the classical gravity approximation is valid. In such cases however, the theory is not actually embeddable in a higher dimensional holographic theory as discussed before. Nevertheless, the presence of UV completion for a large range of irrelevant deformations should not surprise us because the dual quantum theory lives in $0+1$-D.}

We adopt the second method first. It is natural to impose the background metric for the dual quantum theory to be ${\rm d}s^2 = - {\rm d}u^2$. Let us parametrize the cut-off trajectory via the boundary time so that it is given by the functions $z(u)$ and  $t(u)$. The holographic dictionary then implies that the induced metric on the cut-off should be
\begin{equation}
h_{tt}(z(u),t(u)) = - \frac{1}{\epsilon^2}. 
\end{equation}
To achieve this, we will require that 
\be\label{zu}
z(u) = \epsilon t'(u) + \mathcal{O}(\epsilon^2).
\ee
Above the prime denotes differentiation w.r.t. $u$. The function $t(u)$ is determined by the boundary condition on $\Phi$. The key to obtain $SL(2,R)$ symmetry in the IR is to impose the boundary condition where the value of $\Phi$ on the cut-off trajectory satisfies
\begin{equation}\label{Phib}
 \pb (u) = \Phi(z(u), t(u))= \frac{\phi_r(u)}{\epsilon}
\end{equation}
with $\phi_r(u)$ being an arbitrary function which should be specified. In this paper we will set it to be a constant and represent it by $\overline\phi_r$ following \cite{Maldacena:2016upp}. 

We will see later that for well behaved matter sector where we can take the limit $\epsilon \rightarrow 0$, the most singular term in \eqref{phisol1} is indeed $z^{-1}$ and its coefficient is 
 \be
 \alpha_0 (t) = \lim_{\epsilon\rightarrow 0} \alpha_\epsilon (t).
 \ee
Then it follows from \eqref{phisol1}, \eqref{zu} and  \eqref{Phib} that
\begin{align}\label{alpha0}
\alpha_\epsilon(t(u)) &= \overline\phi_r t'(u) + \mathcal{O} (\epsilon^\gamma) \quad {\rm with} \quad \gamma >0, \nonumber\\&\quad {\rm i.e.} \quad \alpha_0(t(u)) = \overline\phi_r t'(u).
\end{align}
For instance, in presence of a minimally coupled free bulk scalar field with $m^2 = 5/16$ the sub-leading term above has $\gamma = 1/2$. 

The dynamics of gravity is then captured by the function $t(u)$ which should be determined from the bulk equations of motion. To see this, we first note that the on-shell action for the pure gravity part is\footnote{One can readily compute the extrinsic curvature $K$ of the cut-off trajectory $\{\gamma(u): (z(u) =\epsilon t'(u), t(u))\}$. The result is that
$$ K =\frac{1 -\epsilon^2\frac{t'''(u)}{t'(u)}}{\left(1 - \epsilon^2 \frac{t''(u)^2}{t'(u)^2}\right)^{\frac{3}{2}}}  =1 - Sch(t,u)\epsilon^2    + \mathcal{O}(\epsilon^4).$$
}
\bea
S^{\rm grav}_{\rm on-shell} &=\frac{1}{8\pi G}\int {\rm d} u \sqrt{-h}\,\,\pb K \,\, = \,\, \frac{1}{8\pi G}\int {\rm d} u \frac{1}{\epsilon} \frac{\overline\phi_r}{\epsilon} K  \nonumber \\ &=
\frac{\overline\phi_r}{8\pi G}\int {\rm d} u \Big(\frac{1}{\epsilon^2}+ {\rm other \,\, singular \,\, terms} \nonumber\\ &- Sch(t,u) + \cdots \Big)
\eea
where
\be
Sch(t,u) = \frac{t'''(u)}{t'(u)} - \frac{3}{2}\frac{t''(u)^2}{t'(u)^2}
\ee
is the Schwarzian derivative. The dotted terms vanish in the limit $\epsilon \rightarrow 0$. The term proportional to $1/\epsilon^2$ and the other singular terms (e.g. one proportional to $\epsilon^{- 3/2}$ which occurs in the presence of a minimally coupled free bulk scalar field with $m^2 = 5/16$) can be subtracted away by appropriate local counterterms (which are built out of the matter sources and their time-derivatives) to render the limit $\epsilon \rightarrow 0$ finite \cite{Cvetic:2016eiv,Gonzalez:2018enk}. We emphasize that new singular terms at subleading orders in $\epsilon$ can appear in the presence of bulk matter. After adding the counterterms and taking the $\epsilon\rightarrow 0$ limit, we obtain
\be\label{onshell-action}
S^{\rm grav}_{\rm on-shell} = \frac{\overline\phi_r}{16\pi G}\int {\rm d} u \,\,(- 2\, Sch(t,u)),
\ee
which gives part of the action for the variable $t(u)$ that determines the cut-off trajectory. 

The matter sector lives in $AdS_2$. The holographic dictionary for this sector can be set up in the traditional way in the limit $\epsilon \rightarrow 0$, although in the presence of a cut-off one needs to set up the dictionary with a bit more care. In this paper we will deal with cases when we can indeed take the limit $\epsilon \rightarrow 0$. The matter sector of course modifies the equation of motion for $t(u)$.

This equation for $t(u)$ can always be obtained from the renormalized on-shell action. However, equivalently assuming that the limit $\epsilon \rightarrow 0$ exists we will be able to also obtain it from the constraint \eqref{alpha-2} rather easily. This will be the topic of our next subsection.

\subsection{Time-reparametrization}\label{tr}
As discussed above, we will deal only with cases in which the limit where the UV cut-off in the dual quantum theory can be taken to infinity or equivalently $\epsilon \rightarrow 0$ exists. In such cases, we can use \eqref{alpha0} which states that $\alpha_0(t(u)) = \overline\phi_r t'(u)$. Then  differentiating both sides of this relation thrice w.r.t. $t$, we readily obtain
\begin{equation}
\dddot\alpha_0 = \overline\phi_r\frac{\left(Sch(t(u) ,u)\right)'}{t'(u)^2}.
\end{equation}
Above dot and prime denote differentiation w.r.t. $t$ and $u$, respectively. From \eqref{alpha-2} it then follows that 
\begin{align}\label{time-repar-gen}
\overline\phi_r  \,\left(Sch(t(u),u)\right)' 
&= t'(u)^2 \lim_{\epsilon\rightarrow 0}\left[T_{zt}(\epsilon, t(u)) -\epsilon\, \partial_t T_{zz}(\epsilon, t(u))\right] .
\end{align}
A necessary condition that the matter sector should satisfy then is that the following limit 
$$ \lim_{\epsilon\rightarrow 0}\left[T_{zt}(\epsilon, t(u)) -\epsilon \dot T_{zz}(\epsilon, t(u))\right]$$
should exist. Of course we should also worry about choosing right integration constants so that we obtain  \eqref{beta-f} and \eqref{alpha-f}.

To see how this can work, we study the example of a minimally coupled free bulk scalar field $\chi$ with $m^2 = 5/16$. The dual operator in the quantum theory has $\Delta = 5/4$. Sourcing the bulk scalar then results in an irrelevant deformation in the dual quantum theory. The Klein Gordon equation
\be
 \partial_z^2\chi -\partial_t^2\chi -\frac{5}{16 z^2} \chi =0
\ee
in the locally $AdS_2$ spacetime has a solution with the following asymptotic expansion
\be\label{asymp1}
\chi(z,t) = J_p(t) z^{-\frac{1}{4}} + O_p(t) z^{\frac{5}{4}}+ \ddot J_p(t) z^{\frac{7}{4}} + \mathcal{O}(z^{\frac{13}{4}}).
\ee
Of course the Klein-Gordon equation can be solved exactly, but at present we will focus only on its asymptotic expansion which is specified completely in terms of $J_p(t)$ and $O_p(t)$. The components of the energy momentum tensor of this field are given by
\bea\label{emcomps}
T_{tt} &=& \frac{1}{2}\left((\partial_t{\chi})^2 + (\partial_z\chi)^2 + \frac{5}{16z^2}\chi^2\right), \nonumber\\
T_{zt} &=& \partial_t{\chi}\partial_z\chi, \nonumber\\
T_{zz} &=& \frac{1}{2}\left((\partial_t{\chi})^2 + (\partial_z\chi)^2 - \frac{5}{16z^2}\chi^2\right).
\eea
Utilizing \eqref{asymp1}, we can readily find that
\begin{align}\label{SchRHS5by4}
\lim_{\epsilon\rightarrow 0}\Big[T_{zt}(\epsilon, t) &-\epsilon \, \partial_t T_{zz}(\epsilon, t)\Big] \nonumber\\
&= \frac{3}{2}\left(\frac{5}{4}O_p(t)\dot J_p(t) +\frac{1}{4}J_p(t)\dot O_p(t) \right)
\end{align}
and therefore we satisfy the necessary condition for our holographic dictionary to make sense in the limit $\epsilon \rightarrow 0$. For this condition to also be sufficient, we need to show that the formal solution of $\Phi$ given by \eqref{phisol1} indeed yields the desired asymptotic behavior in the limit $\epsilon \rightarrow 0$. To examine this, we can again substitute \eqref{asymp1} in \eqref{emcomps} and then in \eqref{phisol1}, and finally take the $\epsilon \rightarrow 0$ limit. This yields
\begin{align}\label{Phiasymp1}
\Phi(z,t) =\lim_{\epsilon \rightarrow 0} &\Big(\frac{\alpha_0(t)}{z} + \frac{J_p^2(t)}{4 \sqrt{z}}\nonumber\\ &+ \left(-\frac{J_p^2(t)}{16 \epsilon^{\frac{3}{2}}} + \beta_\epsilon (t)\right) z + \mathcal{O}(z^{3/2})\Big)
\end{align}
with all other subleading terms not shown here having finite $\epsilon \rightarrow 0$ limit. We observe that the coefficient of  $z$ apparently blows up when $\epsilon \rightarrow 0$ due to presence of a $\epsilon^{- 3/2}$ term. However, utilizing \eqref{beta-f}, we can obtain
\begin{align}\label{beta0int}
\beta_\epsilon(t) &= - C_0 + \frac{1}{8 \epsilon^{\frac{3}{2}}}\int_{-\infty}^t {\rm d}t_1 J_p(t_1) \dot{J_p}(t_1) \nonumber\\
&- \frac{1}{2} \int_{-\infty}^t {\rm d}t_1 \left(\frac{5}{4}O_p(t_1) \dot J_p(t_1)-\frac{1}{4}\dot O_p(t_1) J_p(t_1)\right) + \cdots\nonumber\\
&= - C_0 + \frac{J_p^2(t)}{16 \epsilon^{\frac{3}{2}}}\nonumber\\
&- \frac{1}{2} \int_{-\infty}^t {\rm d}t_1 \left(\frac{5}{4}O_p(t_1) \dot J_p(t_1)-\frac{1}{4}\dot O_p(t_1) J_p(t_1)\right) +\cdots
\end{align}
with $C_0 = \lim_{\epsilon\rightarrow 0} C_\epsilon$ and $\cdots$ indicate terms which vanish in the limit $\epsilon \rightarrow 0$. Crucially we have assumed above that 
\be\label{assume1}
\lim_{t\rightarrow -\infty} J_p^2(t) = 0.
\ee
This assumption is crucial because only with this we get a sensible asymptotic behavior of $\Phi$ in the $\epsilon\rightarrow 0$ limit which can be finally obtained by assembling \eqref{beta-f}, \eqref{alpha-f}, \eqref{SchRHS5by4}, \eqref{Phiasymp1} and \eqref{beta0int}. This asymptotic expansion turns out to be
\begin{align}\label{Phiasymp2}
&\Phi(z,t) = \frac{1}{z}\Bigg(A_0 + B_0 t + C_0 t^2 \nonumber\\ 
&+ \frac{3}{2}\int_{-\infty}^t {\rm d}t_1\int_{-\infty}^{t_1} {\rm d}t_2\int_{-\infty}^{t_2} {\rm d}t_3 \left[\frac{5}{4}O_p(t)\dot J_p(t) +\frac{1}{4}J_p(t)\dot O_p(t) \right]\Bigg) 
\nonumber\\
&+ \frac{J_p^2(t)}{4 \sqrt{z}} - z C_0 \nonumber\\
&- \frac{z}{2} \int_{-\infty}^t {\rm d}t_1 \left(\frac{5}{4}O_p(t_1) \dot J_p(t_1)-\frac{1}{4}\dot O_p(t_1) J_p(t_1)\right)  \nonumber\\&+ \mathcal{O}(z^{\frac{3}{2}}).
\end{align}
In particular, due to \eqref{assume1}, the $\epsilon^{-3/2}$ term in the coefficient of $z$ in \eqref{Phiasymp1} has now been mitigated. An alternative to \eqref{assume1} could have been to set
$$ C_\epsilon = -\lim_{t\rightarrow -\infty}\frac{J_p^2(t)}{16 \epsilon^{\frac{3}{2}}} + \mathcal{O}(\epsilon^0).$$
However this would have implied that $C_0$ is singular. This does not work as $C_0 t^2$ appears in $\alpha_0$ which is the coefficient of the $z^{-1}$ term. We can conclude that if $J_p(t)$ vanishes sufficiently fast in the far past, then the asymptotic expansion of $\Phi$ has non-singular coefficients in the limit $\epsilon\rightarrow 0$. Not only does this guarantee that singular terms in this limit are mitigated but also that the relevant integrals are finite. Eventually $O_p(t)$ is determined from $J_p(t)$ from \textit{regularity} which implements \textit{causal response} in holography. Also clearly because of time-translation symmetry of $AdS_2$, if $J_p(t)$ is constant then so is $O_p(t)$. In this case, although $\Phi$ is modified as evident from \eqref{Phiasymp2}, the $\epsilon\rightarrow 0$ is non-problematic. 

We can thus legitimately investigate the time-reparametrization equation \eqref{time-repar-gen} in the limit $\epsilon\rightarrow 0$ which in our example reduces to 
\begin{align}\label{time-repar-5by4-1}
(Sch(t(u),u))' &= \frac{3 t'(u)^2}{2\overline\phi_r}\Big(\frac{5}{4}O_p(t(u))\dot J_p(t(u))\nonumber\\
&+\frac{1}{4}J_p(t(u))\dot O_p(t(u)) \Big).
\end{align}
The bulk regularity condition which we will study explicitly later implies that
\be
O_p(t) = \int_{-\infty}^t {\rm d} t_1 \,{G_R(t-t_1)} J_p(t_1).
\ee
where $G_R(t-t_1)$ is known. Furthermore, since the boundary time is $u$ as discussed before, the source (perturbation) which couples to the dual operator with $\Delta = 5/4$ is actually
\be\label{Ju}
J(u) = t'(u)^{-\frac{1}{4}} J_p(t(u))
\ee
and similarly the expectation value of the operator that is actually measured as response is
\be\label{Ou}
O(u) = t'(u)^{\frac{5}{4}} O_p(t(u)).
\ee
Therefore,
\be\label{Ouint}
O(u) = \int_{-\infty}^u {\rm d} u_1 \,{G_R(t(u)-t(u_1))} t'(u)^{\frac{5}{4}}t'(u_1)^{\frac{5}{4}}J(u_1).
\ee
Substituting \eqref{Ju} and \eqref{Ou} in \eqref{time-repar-5by4-1}, we obtain
\be\label{time-repar-5by4-2}
(Sch(t(u),u))' = \frac{3}{2\overline\phi_r} \left(\frac{5}{4} O(u) J'(u) + \frac{1}{4} J(u) O'(u)\right).
\ee
The above equation should be understood with $O(u)$ defined via \eqref{Ouint}. Thus the time-reparametrization equation is actually a fourth-order integro-differential equation. It is to be noted that since the Schwarzian is invariant under a fractional linear transformation 
of $t(u)$
$$t(u) \rightarrow \frac{a t(u) +b}{c t(u) + d},$$
as is the reparametrized retarded correlation function
$${G_R(t(u)-t(u_1))} t'(u)^{\frac{5}{4}}t'(u_1)^{\frac{5}{4}},$$ 
owing to the $SL(2,R)$ symmetry of the background $AdS_2$ geometry in which the Klein Gordon equation is solved. We can conclude that the time-reparamtrisation equation retains $SL(2,R)$ symmetry even in the presence of minimally coupled bulk matter.

One can indeed show that in the presence of minimally coupled free bulk scalar field with $-1/4 < m^2 < 3/4$ i.e. corresponding to a deformation with $1/2 <\Delta < 3/2$, the general form of the time-reparametrization equation is
\be\label{time-repar-scalar-gen}
(Sch(t(u),u))' =C_\Delta  \left(\Delta O(u) J'(u) + (\Delta-1) J(u) O'(u)\right).
\ee
with
\be\label{Ouint-gen}
O(u) = \int_{-\infty}^u {\rm d} u_1 \,{G_R(t(u)-t(u_1))} t'(u)^\Delta t'(u_1)^\Delta J(u_1).
\ee
and $C_\Delta = (2\Delta -1)/\overline\phi_r$  which can be set to unity by choosing of $\overline\phi_r = 2\Delta -1$ since only a single scalar deformation is being considered. The equation is symmetric under $SL(2,R)$ transformation of $t(u)$ due to the invariance of
$$ {G_R(t(u)-t(u_1))} t'(u)^\Delta t'(u_1)^\Delta$$
under this transformation. Furthermore, $\Phi$ indeed has an asymptotic expansion with non-singular coefficients in the limit $\epsilon\rightarrow0$.

If $\Delta \geq 3/2$, the leading asymptotic behavior of $\Phi$ is  more singular than $z^{-1}$. For instance, when $\Delta = 3/2$, the matter energy density leads to  leading $z^{-1}\log z$ asymptotics of $\Phi$. The on-shell  action then has $\log \epsilon \,Sch(t,u)$ term which cannot be subtracted by a counterterm which is a local functional of the sources.\footnote{This is similar to the case of a conformal anomaly.} A holographic interpretation of a $\Delta \geq 3/2$ deformation makes sense only after imposing a UV cut-off in the dual theory.

\subsection{A brief tale of three coordinates}
In Jackiw-Teitelboim gravity, the metric is locally always $AdS_2$ and gravity has no local bulk dynamics. Nevertheless, a diffeomorphism of the bulk coordinates which is non-trivial at the boundary has a physical effect as it produces a non-topological on-shell action. This time-reparametrization is described by the variable $t(u)$ which maps the physical (boundary) time $u$ of the observer to the time coordinate $t$ of Fefferman-Graham coordinates. However, due to the $SL(2,R)$ symmetry of the on-shell action and also the equations of motion discussed above, a fractional linear transformation of $t(u)$ has no physical effect on observables such as correlation functions. So the physically distinct solutions of $t(u)$ are members of the \textit{Diff}$/SL(2,R)$ coset.
 
 In absence of matter, the time-reparametrization equation \eqref{time-repar-scalar-gen} implies that the Schwarzian derivative of $t(u)$ must be a constant, i.e.
 \be\label{Schcons}
 Sch(t(u), u) = \pm \frac{2\pi^2}{\beta^2}
 \ee
 with $\beta$ being a real parameter. For the \textit{negative} sign of the Schwarzian derivative of $t(u)$, the solution is
 \be\label{tanh}
 t(u) = \tanh\left(\frac{\pi u}{\beta}\right)
 \ee
up to a $SL(2,R)$ transformation. The three parameters of the $SL(2,R)$ transformation along with $\beta$ supply the necessary four integration constants of \eqref{time-repar-scalar-gen}. If the Schwarzian derivative of $t(u)$ is a \textit{positive} constant, then the solution is 
\be\label{tan}
 t(u) = \tan\left(\frac{\pi u}{\beta}\right)
 \ee
up to a $SL(2,R)$ transformation. In this case, the solution is periodic with period $\beta$. A periodic Lorentzian time does not make sense so we reject such solutions as unphysical. 

For Euclidean signature however, we accept periodic solutions with period $\beta$ as these can indeed be interpreted as thermal solutions with temperature $\beta^{-1}$. Under Euclidean continuation where both $t \rightarrow i t$ and $u \rightarrow i u$, the Schwarzian reverses sign. In this case, only positive constant values of the Schwarzian are physically acceptable. Futhermore, under $u \rightarrow i u$, the Lorentzian solution \eqref{tanh} goes to the Euclidean solution \eqref{tan} such that indeed $t \rightarrow i t $. 

It is natural to ask if we can interpret $t(u)$ in the bulk.  When the cut-off $\epsilon$ is imposed, its trajectory is $z (u)\approx \epsilon t'(u)$ as discussed before. However, when we can take the limit $\epsilon \rightarrow 0$, it is more useful to consider $t(u)$ as the boundary limit of a bulk diffeomorphism. \textit{Of course, the bulk diffeomorphism corresponding to a given $t(u)$ is not unique, so to make such an identification we need gauge fixing.} Instead of retaining Fefferman-Graham gauge where $g_{zz} = 1/z^2$ and $g_{zt} = 0$, we will use ingoing Eddington-Finkelstein gauge in which the $AdS_2$ metric takes the form:
\be\label{EFMu}
{\rm d}s^2 = - \frac{2}{r^2}{\rm d}r {\rm d}u - \left(\frac{1}{r^2}- M(u)\right) {\rm d}u^2
\ee
where $g_{rr} = 0$ and $g_{ru} = -1/r^2$, and the boundary time $u$ is also an ingoing null bulk coordinate. The function $M(u)$ parametrizes the residual gauge freedom, i.e. diffeomorphisms which preserve this gauge. To see this explicitly, we first choose $M(u) = 1$ and write the metric in this gauge as below
\be\label{EFM1}
{\rm d}s^2 = - \frac{2}{\rho^2}{\rm d}\rho {\rm d}\tau - \left(\frac{1}{\rho^2}- 1\right) {\rm d}\tau^2.
\ee
To get back \eqref{EFMu} with an arbitrary $M(u)$ we need to perform the (gauge-preserving) diffeomorphism
\be\label{EFdiffeo}
\tau = \tau (u), \qquad
\rho = \frac{\tau'(u) r}{1- \frac{\tau''(u)}{\tau'(u)} r},
\ee
with
\be\label{SchtauM}
-2 Sch(\tau(u), u) + \tau'(u)^2 = M(u).
\ee
Under such a diffeomorphism, the ingoing null coordinate (observer's boundary time) $u$ maps to $\tau$ which is the ingoing null coordinate (boundary time) of a fixed mass $M =1$ black hole, and this map $\tau(u)$ is determined by the dynamical mass $M(u)$. Furthermore the radial coordinate transforms by a time-dependent fractional linear transformation whose parameters are determined by $\tau(u)$. 

We need to connect the Fefferman-Graham time $t$, with which the bulk metric assumes the \textit{canonical} (Poincare patch) form \eqref{PPform}, to the observer's time $u$. We can do this by first mapping $t$ to $\tau$, and then using the map from $\tau$ to $u$ found above. To bring the bulk metric \eqref{PPform} to the ingoing Eddington-Finkelstein form \eqref{EFM1} with $M(u) = 1$ we need to perform the diffeomorphism
\bea\label{staticdiffeo}
t &=& \frac{1}{2}\left(\tanh\left(\frac{\tau}{2}+ {\rm arctanh}\,\rho\right)+\tanh\left(\frac{\tau}{2}\right)\right), \nonumber\\
z &=& \frac{1}{2}\left(\tanh\left(\frac{\tau}{2}+ {\rm arctanh}\,\rho\right)-\tanh\left(\frac{\tau}{2}\right)\right).
\eea
At the boundary $z =0$, i.e. $\rho = 0$, we find that
\be
t = \tanh\left(\frac{\tau}{2}\right)
\ee
which matches with the form \eqref{tanh} if we set $\beta = 2\pi$. In this case, as follows from \eqref{Schcons}, 
\be\label{Schttau}
Sch(t, \tau) = - \frac{1}{2}.
\ee
In order to obtain the general ingoing Eddington-Finkelstein form of the metric \eqref{EFMu} with an arbitrary $M(u)$ from the canonical Fefferman-Graham coordinates, we simply need to substitute \eqref{EFdiffeo} in \eqref{staticdiffeo}. Then at the boundary $z =0$ i.e. $r = 0$, we find that
\be\label{ttau}
t(u) = \tanh\left(\frac{\tau(u)}{2}\right).
\ee
Utilizing the composition law of the Schwarzian
\be
Sch ( (f\circ g) (u), u) = Sch (g(u), u) + g'(u)^2  Sch ((f\circ g) (u), g(u)),
\ee
\eqref{Schttau} and \eqref{ttau} we find that
\be\label{Sch}
Sch:= Sch(t(u) ,u) = Sch (\tau(u),u) - \frac{1}{2}\tau'(u)^2 
\ee
Comparing with \eqref{SchtauM}, we obtain
\be\label{SchMu}
Sch = - \frac{1}{2}M(u).
\ee
This relates the boundary variable $t(u)$ to the time-dependent black hole mass $M(u)$, and thus provides a bulk interpretation of $t(u)$. The actual ADM mass of the black hole is \cite{Almheiri:2014cka,Maldacena:2016upp}
\be
M_{\rm ADM}(u) =\frac{\overline\phi_r}{16\pi G} (- 2\, Sch) = \frac{\overline\phi_r}{16\pi G} M(u).
\ee
Therefore,
\be\label{relation}
-Sch(t(u),u) = \frac{8 \pi G}{\overline\phi_r} M_{\rm ADM}(u).
\ee
\Ayan{The pure JT on-shell gravitational action \eqref{onshell-action} in the presence of a minimally coupled bulk scalar field is modified to}
\begin{align}\label{onshell-full}
S^{\rm grav}_{\rm on-shell} &= \frac{\overline{\phi}_r}{16\pi G}\int {\rm d}u\,(- 2 Sch(t(u),u)) \nonumber\\
&+ \frac{1}{16\pi G} \int {\rm d}u \, J(u) O (u).
\end{align}
{\Ayan{One can also readily derive the equation of motion for $t(u)$ given by \eqref{time-repar-scalar-gen} from the above action after expressing $M_{\rm ADM}$ in terms of the Schwarzian of $t(u)$, and also $O(u)$ in terms of $J(u)$ and $t(u)$ via \eqref{Ouint-gen} \cite{Maldacena:2016upp}. We have derived this time-reparametization equation in the previous subsection from the bulk gravitational constraints instead.}
\subsection{The second law and the profile of the dilaton}\label{dilpf}
Here we review results presented in \cite{Duchting:2000qk} regarding the second law in JT gravity, and then we discuss how to obtain the profile of the dilaton in the physical geometry corresponding to the observer's time at the boundary which in the ingoing Eddington-Finkelstein gauge takes the form of \eqref{EFMu}. 

Since in $1+1$-D spacetime any smooth null curve is a null geodesic, for any smooth null curve $x^\mu(\lambda)$ we can find an affine parmeter $\lambda$ such that $l^\mu = {\rm d}x^\mu/d\lambda$ satisfies $l^\mu l_\mu = 0$ and
\be
(l\cdot\nabla) l^\mu = 0.
\ee
Contracting the dilaton equation \eqref{PhiCons} we obtain
\be
(l \cdot \nabla)^2 \Phi = - T_{\mu\nu} l^\mu l^\nu
\ee
i.e.
\be
\frac{{\rm D}^2 \Phi}{D\lambda^2} = - T_{\mu\nu} l^\mu l^\nu.
\ee
Assuming
\be\label{assume-ent}
\frac{{\rm d}\Phi}{{\rm d}\lambda} \bigg\vert_{\lambda \rightarrow  \infty} = 0,
\ee
it follows that
\be
\frac{{\rm d}\Phi}{{\rm d}\lambda} = \int_\lambda^\infty {\rm d}\lambda_1 T_{\mu\nu} l^\mu l^\nu.
 \ee
Since classical bulk matter satisfies the null energy condition $T_{\mu\nu} l^\mu l^\nu > 0$, we obtain that
\be
\frac{{\rm d}\Phi}{{\rm d}\lambda} > 0.
\ee
Therefore $\Phi (\lambda)$ is a monotonically increasing function on a null curve where \eqref{assume-ent} is satisfied\footnote{It is a different question of course if such a monotonically increasing entropy function also satisfies a first law. We will not deal with this issue here.}. Such a null curve can be readily found if the full geometry settles down to a static configuration at late time -- it is the one which coincides with the apparent horizon $r_h(u) = 1/\sqrt{M(u)}$ at late time in the coordinates where the metric assumes the form \eqref{EFMu}. If the geometry does not become static and/or the limit \eqref{assume-ent} does not exist because the null geodesic cannot be extended to arbitrarily large affine time in the future, then a second law need not hold in the classical solution. 

The equation determining the dilaton profile \eqref{PhiCons} takes a much simpler form in the coordinates \eqref{EFMu} than what we obtained before in the case of Fefferman-Graham. In presence of a free scalar field $\chi$ with $m^2 = \Delta(\Delta -1)$ and minimally coupled to the metric, the $rr-$compoment of \eqref{PhiCons} is simply
\begin{equation}\label{Phirreq}
\partial_r^2 \Phi + 2 \frac{\partial_r \Phi}{r} = - (\partial_r \chi)^2.
\end{equation}
The Klein-Gordon equation for $\chi$ is
\begin{equation}\label{KGEFMu}
\partial_r (d_+ \chi)  + \frac{\Delta(\Delta -1)}{2 r^2}\chi = 0,
\end{equation}
where
\begin{equation}\label{chieqEF}
d_+ \chi = \partial_u \chi - \frac{1}{2}(1  - r^2 M(u)) \partial_r\chi.
\end{equation}
We note that $d_+ \chi \equiv (\xi \cdot \nabla)\chi$, i.e. the directional derivative of $\chi$ along the outgoing null direction $\xi^\mu$ with $\xi^u = 1$ and $\xi^r = - (1/2)(1- r^2 M(u))$.  

We should choose $1 < \Delta < 3/2$. For concreteness, let $\Delta =  5/4$. The asymptotic expansion of $\chi$ which follows from \eqref{chieqEF} is
\begin{eqnarray}
\chi & = & J(u) r^{-\frac{1}{4}} + J'(u) r^{\frac{3}{4}} + O(u) r^{\frac{5}{4}}
\nonumber\\&& +\left(\frac{3}{2}J''(u) - \frac{3}{16}J(u) M(u)\right)r^{\frac{7}{4}}+ O'(u) r^{\frac{9}{4}} + \cdots.
\end{eqnarray}
Then the solution of \eqref{Phirreq} is
\begin{align}\label{PhisolEF}
\Phi  &= \frac{a(u)}{r}+ b(u) + \frac{J(u)^2}{4}r^{- \frac{1}{2}}\nonumber\\&
+ \int_0^r {\rm d}r'' \frac{1}{r''^2} \Big(\frac{1}{8}J(u)^2 r''^{\frac{1}{2}}\nonumber\\
&\qquad\qquad\qquad-\int_0^{r''} {\rm d}r' r'^2 \left( \partial_{r'}\chi(r', u) \right)^2  \Big).
\end{align}
Note both the integrals above are finite. One can of course write similar expressions for arbitrary $\Delta$.

The $rr-$ and $tt-$components of  \eqref{PhiCons} are of course constraints and therefore they reduce to equations determining $a(u)$ and $b(u)$ in \eqref{PhisolEF}. These are
\begin{align}\label{PhiconsEFa}
&b(u) = a'(u),\\ \label{PhiconsEFb}
&a'''(u) - M(u) a'(u) - \frac{1}{2} a(u) M'(u) \nonumber\\ &\qquad= (2\Delta -1)\left(\Delta O(u) J'(u) + (\Delta -1) O'(u) J(u)\right).
\end{align}

We now observe that since the boundary time corresponds to the observer's time in the $r$ and $u$ coordinates, the Dirichlet boundary condition for $\Phi$ simply implies that
\begin{equation}\label{bcEF}
a(u) = \overline{\phi}_r.
\end{equation}
We choose $\overline{\phi}_r$ as a constant. Substituting this in \eqref{PhiconsEFb} we obtain that
\begin{align}
b(u) &= 0, \label{Phicons2a}\\
- \frac{1}{2}  M'(u) &= C_\Delta\left( \Delta O(u) J'(u) + (\Delta -1) O'(u) J(u)\right), \nonumber\\
\quad \text{with  } C_\Delta \, &=\, \frac{2\Delta -1}{\overline{\phi}_r }\label{Phicons2b}.
\end{align}
We readily note that the content of the above equation is nothing but the time-parametrization $t(u)$, which characterizes the map from the physical time $u$ to the time $t$ of the vacuum. To see this, we recall our result from the previous subsection that the change of coordinates which takes our present  metric \eqref{EFMu} to pure $AdS_2$ with $M(u) = 0$ implies a time-reparametrization at the boundary $t(u)$ given by $Sch(t(u),u) = - (1/2) M(u)$. Substituting this in \eqref{Phicons2b} we indeed recover our old time-reparametrization equation \eqref{time-repar-scalar-gen}. We also recall that we choose $\overline{\phi}_r = 2 \Delta -1$, so that $C_\Delta = 1$. 

The combination of boundary condition on the dilaton \eqref{bcEF} and the constraint \eqref{PhiconsEFb} thus determines the mass $M(u)$ in the Eddington-Finkelstein coordinates even when $\overline{\phi}_r$ is not a constant. Then we need to use the relation $M(u) = - 2 Sch(t(u),u)$ here to obtain the time-reparametrization equation with a general time-dependent $\overline{\phi}_r$.

The dilation profile, e.g. for $\Delta = 5/4$, then can be obtained from \eqref{PhisolEF} and it is
\begin{align}\label{PhisolEFfin}
\Phi  &= \frac{3}{2 r}+ \frac{J(u)^2}{4}r^{- \frac{1}{2}}\nonumber\\
&+\int_0^r {\rm d}r'' \frac{1}{r''^2} \Big( \frac{1}{8}J(u)^2 r''^{\frac{1}{2}}\nonumber\\
&\qquad \qquad\qquad-\int_0^{r''} {\rm d}r' r'^2 \left( \partial_{r'}\chi(r', u) \right)^2  \Big),
\end{align}
where we have set $\overline{\phi}_r = 2 \Delta -1 = 3/2$. This explicit form will help us to determine whether the second law can be indeed satisfied. 

\section{Finding explicit time-dependent solutions}
\subsection{Conserved charges and Ward identities}
In the case of pure JT gravity, the Noether charges corresponding to the $SL(2,R)$ symmetries have been discussed in \cite{,Maldacena:2016upp}. The infinitesimal $SL(2,R)$ transformations are $t(u) \rightarrow t(u) + \epsilon\, \delta t(u)$, with $\delta t(u)  = {1, t(u), t(u)^2}$ generating translation, dilation and special conformal transformation respectively. The corresponding conserved charges are:
\begin{align}
Q_0 &= \frac{t'''(u)}{t'(u)^2}-\frac{t''(u)^2}{t'(u)^3}, \\
Q_1 &= t(u)\left(\frac{t'''(u)}{t'(u)^2}-\frac{t''(u)^2}{t'(u)^3}\right) - \frac{t''(u)}{t'(u)},\\
Q_2 &=t(u)^2\left(\frac{t'''(u)}{t'(u)^2}-\frac{t''(u)^2}{t'(u)^3}\right) -2t (u)\left(\frac{t''(u)}{t'(u)}-\frac{t'(u)}{t(u)}\right).
\end{align}
We can readily see that
\be
Q_i'(u) = \frac{t(u)^i}{t'(u)} Sch' 
\ee
for $i = 0,1,2$ so that indeed these are conserved on-shell in pure Teitelboim-Jackiw gravity, i.e. when  $Sch(t(u),u)$ is a constant.  Furthermore, the Casimir
\be
Q_1^2 - Q_0 Q_2 = - 2 \,Sch
\ee
is a constant in the absence of matter. For later convenience, we define the Noether charges
\bea
Q = \frac{1}{2}(Q_0 -Q_2), \quad Q_\pm = \frac{1}{2}(Q_0 +Q_2 \pm 2 Q_1).
\eea
Shifting to the variable $\tau (u)$, which is the boundary time of the $M(u) =1$ black hole and is related to $t(u)$ via \eqref{ttau}, we obtain the explicit forms
\bea\label{3Qs}
Q &=& \frac{\tau'''(u)}{\tau'(u)^2}-\frac{\tau''(u)^2}{\tau'(u)^3}-\tau'(u), \\
Q_+ &=&\left(\frac{\tau'''(u)}{\tau'(u)^2}-\frac{\tau''(u)^2}{\tau'(u)^3}-\frac{\tau''(u)}{\tau'(u)}\right)e^{\tau(u)},\\
Q_- &=&\left(\frac{\tau'''(u)}{\tau'(u)^2}-\frac{\tau''(u)^2}{\tau'(u)^3}+\frac{\tau''(u)}{\tau'(u)}\right)e^{-\tau(u)},
\eea
which satisfy
\bea\label{WIbasic}
Q' = \frac{1}{\tau'(u)} Sch', \quad Q_\pm' = \frac{e^{\pm \tau(u)}}{\tau'(u)} Sch'.
\eea
Furthermore, the Casimir is
\be
Q^2 - Q_+Q_- = - 2 \, Sch.
\ee
We note that all derivatives of $\tau$ at a given value of $\tau$ can be expressed in terms of the Noether charges:
\small
\begin{align}\label{iden1}
\tau'&= \frac{1}{2}\left(Q_- e^{\tau} + Q_+ e^{-\tau}- 2Q\right), \\\label{iden2}
\tau''&= \frac{1}{4}\left(Q_- e^{\tau} - Q_+ e^{-\tau}\right)\left(Q_- e^{\tau} + Q_+ e^{-\tau}- 2Q\right),\\\label{iden3}
\tau'''&= \frac{1}{4}\left(Q_-^2 e^{2\tau} +Q_+^2e^{-2\tau}-Q\left(Q_- e^{\tau} + Q_+ e^{-\tau}\right)\right)\nonumber\\
&\times\left(Q_- e^{\tau} + Q_+ e^{-\tau}- 2Q\right).
\end{align}
\normalsize
One can then take the following approach to obtain all solutions of $\tau(u)$ in the absence of matter. At the initial moment $u = u_{in}$, we need to specify the value of $\tau(u_{in})$ and the three Noether charges. \textit{In the absence of matter, the values of these Noether charges do not change.} At the initial instant we can then use \eqref{iden1} to obtain $\tau'(u_{in})$.  Next, we update $\tau$ using
$$\tau (u_{in}+ \Delta u) =\tau(u_{in}) + \tau'(u_{in})\Delta u.$$
Since we have $\tau(u_{in}+ \Delta u)$, we can use \eqref{iden1} again to obtain $\tau'(u_{in}+ \Delta u)$. We can thus continue further to generate $\tau(u)$ from the initial data given by $\tau(u_{in})$ and the three (constant) values of the Noether charges.

Note that one can always set the Noether charges to the following constant values
\be\label{Qcan}
Q = -\frac{2\pi}{\beta}, \quad Q_\pm = 0
\ee
via an appropriate $SL(2,R)$ transformation. In this case, $Sch = -2\pi^2/\beta^2$ and
\be
\tau(u)=\tau(u_{in})+ \frac{2\pi}{\beta}(u- u_{in}).
\ee
Furthermore, without changing the values of the charges given by \eqref{Qcan}, we can set
$$\tau(u_{in}) =   \frac{2\pi}{\beta} u_{in}$$
and reproduce \eqref{tanh}. Setting the value of $\tau(u_{in})$ also amounts to a $SL(2,R)$ transformation of $t(u)$. Actually for any choice of Noether charges there will be a one parameter family of $SL(2,R)$ transformations which will leave them invariant -- this family then defines the chosen $SL(2,R)$ \textit{frame}. A $SL(2,R)$ transformation has no effect on physical observables, therefore we can derive all real-time properties of thermal equilibrium state at temperature $\beta^{-1}$ from this simple solution \eqref{tanh} which is linear in $u$.

For a constant value of $Sch = - 2\pi^2/\beta^2$, we can parametrize all \textit{real} values of $SL(2,R)$ charges as follows
\begin{align}\label{Qpar}
& Q = - \frac{2\pi}{\beta}\cosh\theta\cos\phi, \quad Q_- = \frac{2\pi}{\beta}(\sinh\theta\cos\phi + \sin\phi),\nonumber\\
& Q_+ =  \frac{2\pi}{\beta}(\sinh\theta\cos\phi - \sin\phi).
\end{align}
The general solution corresponding to the above charges are:
\begin{widetext}
\be\label{taupar}
\tau(u) =\frac{\beta}{\pi} {\rm arctanh}\left(\frac{e^{\frac{\theta}{2}}\left(\cosh \frac{\eta}{2}\cos \frac{\phi}{2}+\sinh \frac{\eta}{2}\sin \frac{\phi}{2}\right)\tanh\left(\frac{\pi}{\beta}u\right)+e^{\frac{\theta}{2}}\left(\sinh \frac{\eta}{2}\cos\frac{\phi}{2}+\cosh \frac{\eta}{2}\sin \frac{\phi}{2}\right)}{e^{-\frac{\theta}{2}}\left(\sinh \frac{\eta}{2}\cos \frac{\phi}{2}-\cosh \frac{\eta}{2}\sin \frac{\phi}{2}\right)\tanh\left(\frac{\pi}{\beta}u\right)+e^{-\frac{\theta}{2}}\left(\cosh \frac{\eta}{2}\cos\frac{\phi}{2}-\sinh \frac{\eta}{2}\sin \frac{\phi}{2}\right)}\right).
\ee
\end{widetext}
\normalsize
The parameters $\theta$, $\phi$ and $\eta$ represent an $SL(2,R)$ transformation of $t(u)$ as should be clear from \eqref{tanh}. However, it is explicit in \eqref{Qpar} that only $\theta$ and $\phi$ along with $\beta$ determine the Noether charges. The parameter $\eta$ nevertheless sets the value of $\tau(u_{in})$ and is thus \textit{not} a redundant variable.  The above parametrization will be useful in characterising the dynamics in the presence of matter.

\subsection{The algorithm}\label{algo}
When bulk matter satisfies appropriate conditions, the $SL(2,R)$ symmetry of the time-reparametrization equation \eqref{time-repar-scalar-gen} is preserved. Nevertheless, the modified Noether charges are not local. Therefore, it is more useful to derive the modified Ward identities of the Noether charges of the pure Schwarzian action which can be obtained from \eqref{time-repar-scalar-gen} and \eqref{WIbasic}. Setting $C_\Delta = 1$ by choosing $\overline{\phi}_r$ appropriately, these modified Ward identities are
\small
\begin{align}\label{WI1}
Q' &= \tau'(u)\Big(\Delta O_{th}(\tau(u)) \frac{{\rm d}J_{th}(\tau(u))}{{\rm d}\tau (u)}\nonumber\\
&\quad+(\Delta -1) J_{th}(\tau(u)) \frac{{\rm d}O_{th}(\tau(u))}{{\rm d}\tau (u)}\Big), \\\label{WI2}
Q_+' &= e^{\tau(u)}\tau'(u)\Big(\Delta O_{th}(\tau(u)) \frac{{\rm d}J_{th}(\tau(u))}{{\rm d}\tau (u)}\nonumber\\
&\quad+(\Delta -1) J_{th}(\tau(u)) \frac{{\rm d}O_{th}(\tau(u))}{{\rm d}\tau (u)}\Big),
\\\label{WI3}
Q_-' &= e^{-\tau(u)}\tau'(u)\Big(\Delta O_{th}(\tau(u)) \frac{{\rm d}J_{th}(\tau(u))}{{\rm d}\tau (u)}\nonumber\\
&\quad+(\Delta -1) J_{th}(\tau(u)) \frac{{\rm d}O_{th}(\tau(u))}{{\rm d}\tau (u)}\Big),
\end{align}
\normalsize
where
\bea\label{Jth}
&J_{th}(\tau(u)) = J(u) \tau'(u)^{\Delta -1}, \\\label{Oth}
 &O_{th}(\tau(u)) = O(u) \tau'(u)^{-\Delta}.
\eea
The time-reparametrization equation \eqref{time-repar-scalar-gen} itself can be written in the form
\be\label{WIH}
\frac{{\rm d}H(u)}{{\rm d}u} = J'(u) O(u)
\ee
where we can readily identify $H(u)$ with the Hamiltonian, i.e. the Noether charge corresponding to the $u$-translation symmetry which is broken explicitly in the presence of $J(u)$. This Hamiltonian explicitly is
\begin{align}
H(u) &= Sch(t(u), u) - (\Delta - 1) J(u) O(u) \nonumber\\
&= Sch(\tau(u), u) - \frac{1}{2}\tau'(u)^2 \nonumber\\
&-  (\Delta - 1) \tau'(u) J_{th}(\tau(u)) O_{th}(\tau(u)).
\end{align}
\normalsize
Above, we have used \eqref{Sch}. It is to be noted that each of the four Ward identities, namely \eqref{WI1}, \eqref{WI2}, \eqref{WI3} and \eqref{WIH} implies the equation of motion \eqref{time-repar-scalar-gen} for $\tau(u)$ and has no content otherwise. Nevertheless, we will be able generate time-dependent solutions via exploiting the integral forms of the three Ward identities \eqref{WI1}, \eqref{WI2} and \eqref{WI3}. The Ward identity \eqref{WIH} will provide a consistency check and accuracy test for numerics.

In order to proceed further, we will need to understand how to obtain $O(u)$ self-consistently from $J(u)$. If we know $t(u)$ for $u< u_0$, then \eqref{Ouint-gen} tells us how to obtain $O(u)$. The problem is that the integral in  \eqref{Ouint-gen} can only be defined via an appropriate analytic continuation for which it is necessary to first go to frequency space -- this will be a cumbersome procedure for a non-trivial $t(u)$ which is not linear or a simple function of $u$. 

This difficulty can be readily circumvented via $J_{th}(\tau(u))$ and $O_{th}(\tau(u))$ defined in \eqref{Jth} and \eqref{Oth}. These are the scalar source and response respectively corresponding to the bulk scalar field $\chi$ living in the metric \eqref{EFM1} with $M(u) = 1$ and with boundary time $\tau(u)$ as discussed before. In these coordinates, the form of the Klein-Gordon equation is simply a special case of that given by \eqref{KGEFMu} and \eqref{chieqEF} with $M(\tau) =1$, i.e. 
\be
\partial_\rho(d_+\chi) + \frac{\Delta(\Delta-1)}{2\rho^2}\chi = 0
\ee
where
\be
d_+ = \xi\cdot\nabla, \quad {\rm with} \quad \xi^\rho =- \frac{1}{2}(1-\rho^2), \,\xi^\tau = 1.
\ee
With an input of $J_{th}(\tau)$ obtained from \eqref{Jth} and initial conditions $\chi(\rho, \tau = 0)$, we can readily solve this equation via the method of characteristics to obtain $O_{th}(\tau)$. From the latter, we can extract $O(u)$ if needed utilizing \eqref{Oth}.

To see how this works explicitly, we take the specific case of $\Delta = 5/4$. It is useful to first define
\be
\overline{d_+\chi} := d_+\chi - \frac{1}{8}J_{th}(\tau) \rho^{-\frac{5}{4}}- \frac{5}{8}\frac{{\rm d}J_{th}(\tau)}{\rm d \tau} \rho^{-\frac{1}{4}}
\ee
because $\overline{d_+\chi}$ has a non-singular asymptotic expansion 
\be\label{asymp-Oth}
\overline{d_+\chi} \approx - \frac{5}{8}O_{th}(\tau) \rho^{\frac{1}{4}}
\ee
near $\rho = 0$. We note that
\be\label{chi-dot}
\partial_\tau \chi = \overline{d_+\chi} +\frac{1}{2}(1-\rho^2)\partial_\rho\chi+ \frac{1}{8}J_{th}(\tau) \rho^{-\frac{5}{4}}+ \frac{5}{8}\frac{{\rm d}J_{th}(\tau)}{\rm d \tau} \rho^{-\frac{1}{4}}.
\ee
Furthermore, the equation of motion for $\overline{d_+\chi}$ is
\be
 \partial_\rho\overline{d_+\chi}+ \frac{5}{32\rho^2}\left(\chi - J_{th}(\tau) \rho^{-\frac{1}{4}}-\frac{{\rm d}J_{th}(\tau)}{\rm d \tau} \rho^{\frac{3}{4}}\right) = 0.
\ee
so that
\begin{align}\label{dpluschi}
\overline{d_+\chi}(\rho, \tau) &= - \int_o^\rho {\rm d}\rho_1\,\, \frac{5}{32\rho_1^2}\Big(\chi(\rho_1, \tau) - J_{th}(\tau) \rho_1^{-\frac{1}{4}}\nonumber\\
&\qquad\qquad\qquad\qquad-\frac{{\rm d}J_{th}(\tau)}{\rm d \tau} \rho_1^{\frac{3}{4}}\Big).
\end{align}
Crucially the integral above on the right hand side is finite. Therefore, if we are given an initial profile $\chi(\rho, \tau = \tau_{in})$ and we also know $J_{th}(\tau)$ for all $\tau < \tau_0$, we can readily obtain $O_{th}(\tau)$ for all  $\tau < \tau_0$ as follows. First, given $\chi(\rho, \tau = \tau_{in})$ at initial time, we can use \eqref{dpluschi} to generate $\overline{d_+\chi}$. From the latter, we can obtain $\partial_\tau\chi$ at initial time utilizing \eqref{chi-dot}. The knowledge of $\partial_\tau\chi$ then allows us to propagate $\chi$ to the next time instant. We can thus continue and generate $\chi(\rho, \tau)$ along with $\overline{d_+\chi}(\rho, \tau)$ up to the instant we know $J_{th}(\tau)$ exactly. Furthermore, the asymptotic expansion \eqref{asymp-Oth} allows us to extract $O_{th}(\tau)$.

We are now ready to describe our algorithm for finding $\tau(u)$ for a given $J(u)$. This relies primarily on the integrated form of the following Ward identities \eqref{WI1}, \eqref{WI2} and \eqref{WI3}:
\begin{widetext}
\bea\label{WI1-int}
Q(u) - Q(u_{in}) &=& \int_{\tau(u_{in})}^{\tau(u)}{\rm d}\tau_1 \,\left(\Delta O_{th}(\tau_1) \frac{{\rm d}J_{th}(\tau_1)}{{\rm d}\tau_1}+(\Delta -1) J_{th}(\tau_1) \frac{{\rm d}O_{th}(\tau_1)}{{\rm d}\tau_1}\right), \\\label{WI2-int}
Q_+(u) - Q_+(u_{in}) &=& \int_{\tau(u_{in})}^{\tau(u)}{\rm d}\tau_1\, e^{\tau_1}\left(\Delta O_{th}(\tau_1) \frac{{\rm d}J_{th}(\tau_1)}{{\rm d}\tau_1}+(\Delta -1) J_{th}(\tau_1) \frac{{\rm d}O_{th}(\tau_1)}{{\rm d}\tau_1}\right),
\\\label{WI3-int}
Q_-(u) - Q_-(u_{in}) &=& \int_{\tau(u_{in})}^{\tau(u)}{\rm d}\tau_1 \,e^{-\tau_1}\left(\Delta O_{th}(\tau_1) \frac{{\rm d}J_{th}(\tau_1)}{{\rm d}\tau_1}+(\Delta -1) J_{th}(\tau_1) \frac{{\rm d}O_{th}(\tau_1)}{{\rm d}\tau_1}\right).
\eea
\end{widetext}
\normalsize
Our algorithm then consists of the following steps
\begin{enumerate}
\item Given initial values of $\tau(u_{in})$ and the three $SL(2,R)$ charges, we can extract $\tau'(u_{in})$ using \eqref{iden1} and $\tau''(u_{in})$ using \eqref{iden2}.
\item From $\tau'(u_{in})$ and known $J(u)$, we can extract $J_{th}(\tau(u_{in}))$ using \eqref{Jth} and then ${\rm d}J_{th}/{\rm d}\tau$ at $\tau(u_{in})$ since we also know $\tau''(u_{in})$.
\item Given initial profile of $\chi$ (more on this later), $J_{th}$ and ${\rm d}J_{th}/{\rm d}\tau$ at $\tau(u_{in})$ we extract the initial profile of $\overline{d_+\chi}$.
\item We then obtain $O_{th}(\tau)$ at $\tau(u_{in})$ using \eqref{asymp-Oth}. 
\item We can now update the three $SL(2,R)$ charges corresponding to the next time instant using \eqref{WI1-int}, \eqref{WI2-int}, \eqref{WI3-int}.
\item We propagate $\tau$ to the next time instant using $\tau'(u_{in})$. Furthermore, we propagate the radial profile of $\chi$ to the next time instant utilizing $\partial_\tau\chi$ which can be extracted from known $\overline{d_+\chi}$ via \eqref{chi-dot}.
\item We repeat all steps above at the next time instant.
\end{enumerate}
It is to be noted that we are always evolving the bulk scalar field in a geometry whose boundary time is $\tau(u)$ and not $u$ itself and corresponding to $M(u) = 1$ black hole. This however requires constant remapping of the source and also the response as discussed above. The integrated form of the Ward identity \eqref{WIH} 
\be
H(u) - H(u_{in}) = \int_{u_{in}}^u {\rm d}u_1 \, J'(u_1) O(u_1)
 \ee
 can be used to check the accuracy of the numerics. For doing this, we will need to extract $O(u)$ from $O_{th}(\tau(u))$ using \eqref{Oth}.

Instead of specifying the initial values of the three Noether charges along with $\tau(u_{in})$, we could have provided $\tau'(u_{in})$, $\tau''(u_{in})$ and $\tau'''(u_{in})$. Then one can use our algorithm by initializing the values of the Noether charges via \eqref{3Qs}. It is more physical to provide the initial values of the Noether charges though. 

The presence of a source $J(u)$ essentially has two physical effects: (i) it makes the Hamiltonian $H$ time-dependent, and (ii) it also varies the $SL(2,R)$ frame along with the Hamiltonian by making all Noether charges \eqref{3Qs} time-dependent and implying that even if the system settles down in the far future with a constant value of the Hamiltonian, the $SL(2,R)$ frame will still be generically different from the initial one. The difference between initial and final $SL(2,R)$ frames can be detected via long-time correlations between far past and far future. 

However, only the relative difference between the initial and final $SL(2,R)$ frames is physical because a time-independent $SL(2,R)$ transformation of the full solution (which also changes initial data) will surely have no physical effect. 

At any point of time, since we know the three Noether charges we can use \eqref{Qpar} to obtain the instantaneous values of the three parameters $\beta(u)$, $\theta(u)$ and $\phi(u)$. Substituting these instantaneous values in \eqref{taupar} and matching the right hand side with $\tau(u_{in})$ at $u= u_{in}$, we can determine the instantaneous $\eta(u)$ as well. It is to be noted that we are not promoting $\beta$, $\theta$, $\phi$ and $\eta$ to time-dependent variables in \eqref{taupar}.\footnote{If we do so, then $\eta$ will be a fixed constant determined by $\tau(u_{in})$ and the initial values of the three Noether charges, and not a time-dependent variable.} Rather we are matching with this form at every instant independently to extract the instantaneous values of the four parameters $\beta(u)$, $\theta(u)$, $\phi(u)$ and $\eta(u)$. Representing the instantaneous functional form of $\tau(u)$ via these four parameters helps us to track the change in $SL(2,R)$ frame of the pure $AdS_2$ boundary time $t(u)$, which is encoded by $\theta(u)$ and $\phi(u)$, along with $\eta(u)$ and the value of the $SL(2,R)$ invariant $Sch$ which is given by $Sch = -2\pi^2/\beta(u)^2$.

The class of problems we will examine in the next subsection will correspond to perturbing a pre-existing thermal state by a scalar source which decays sufficiently fast in time. In this case, the minimally coupled bulk scalar fields will vanish initially in absence of sources as otherwise they will have singular profiles. Therefore, we will choose initial conditions where $\chi$ vanishes on the initial time surface -- if chosen sufficiently far in the past, then it will be so in any bulk coordinate system. Furthermore, due to the presence of $SL(2,R)$ symmetry, we can always set initial temperature to be $1/(2\pi)$ (i.e. $\beta(u\rightarrow-\infty) = 2\pi$ and $M(u\rightarrow-\infty) =1$ in the bulk) and furthermore $\tau(u\rightarrow-\infty) \approx u$ can be set initially by an appropriate time-independent $SL(2,R)$ transformation as discussed before. This will also imply that if $u_{in}$ is in the far past we can choose
\be\label{Q-init}
\tau(u_{\rm in}) = u_{\rm in}, \quad Q(u_{\rm in}) = -1, \quad Q_+(u_{\rm in}) = Q_-(u_{\rm in}) = 0.
\ee

\paragraph{An alternative algorithm:} The reader has possibly already noted that we could have followed an alternative route where we need not have conformally mapped the source $J(u)$ to that of a state with a constant temperature. In this case, we could have used the bulk geometry \eqref{EFMu} updating $M(u) = - 2 Sch = Q^2 - Q_+ Q_-$ along with the $SL(2,R)$ charges. The Klein-Gordon equation in this geometry given by \eqref{KGEFMu} and \eqref{chieqEF} features only $M(u)$ but not its derivative. Therefore, we can solve the Klein-Gordon equation in these coordinates with the physical source $J(u)$ and thus obtain $O(u)$ directly via the method of characteristics. However, it turns out that especially in the semi-holographic case, $J_{th}$ and $O_{th}$ give us useful insights. The alternative algorithm however is useful for obtaining the profile of the bulk dilaton utilizing \eqref{PhisolEFfin}. This alternative algorithm also serves the purpose of cross-checking numerical results.

\subsection{A typical pumped state in $NAdS_2$ holography}
We study the typical case of a Gaussian source $J(u)$ which couples to an operator $O(u)$ with $\Delta = 5/4$ following the algorithm mentioned before. As mentioned, we choose the mass of the pre-existing black hole to have unit mass and without loss of generality the standard $SL(2,R)$ frame where only $Q$ is non-zero.

\begin{figure}
\begin{center}
\begin{center}
\subfloat[Plot of $J(u)$.\label{Fig:Ju}]
{ \includegraphics[width=0.5\linewidth]{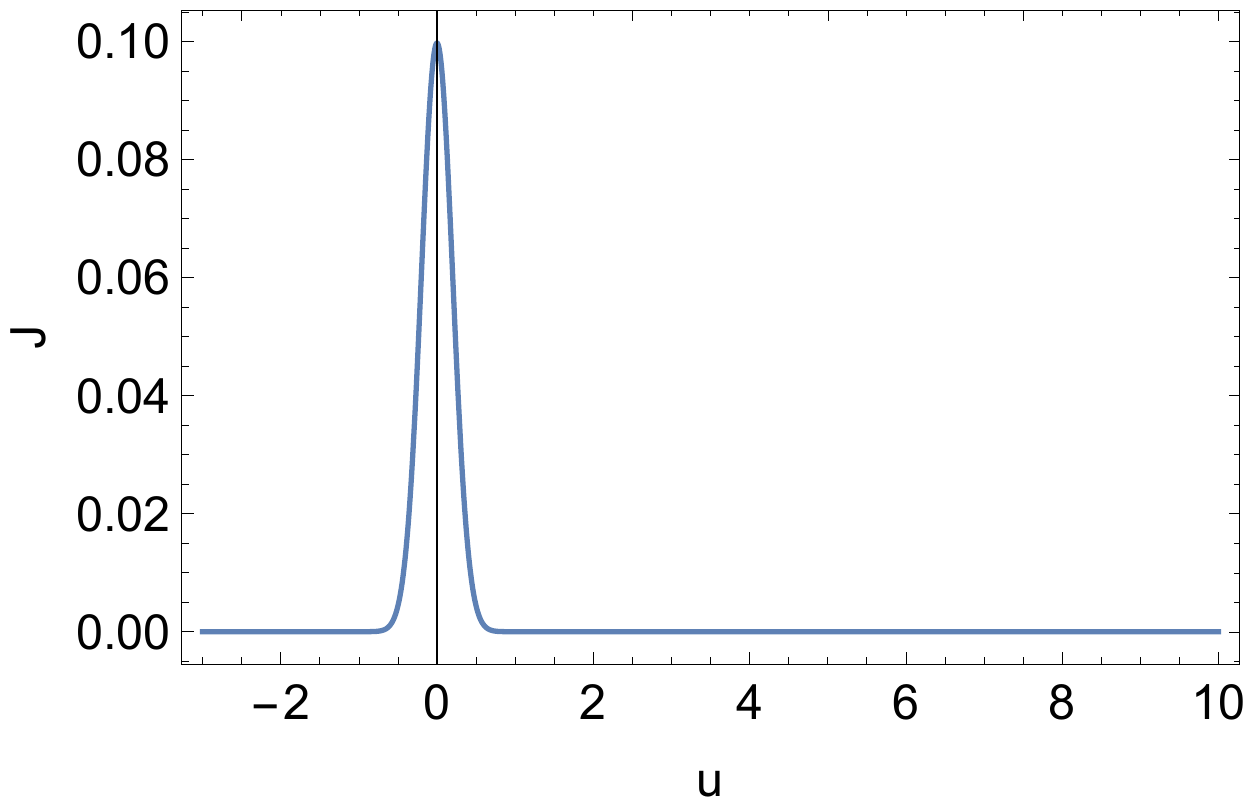}} 
\subfloat[Plot of $O(u)$.\label{Fig:Ou}]
{ \includegraphics[width=0.5\linewidth]{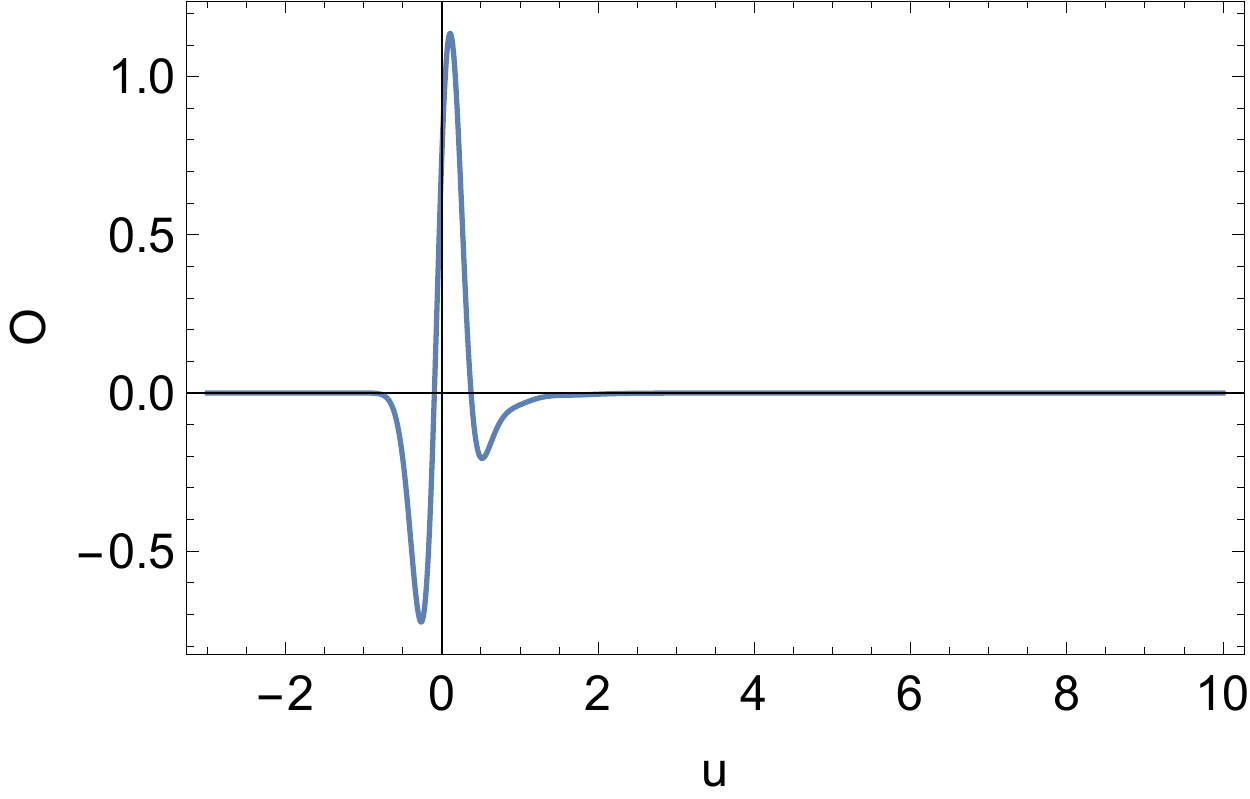}}
\noindent \\
\subfloat[Plot of $J_{th}(\tau(u))$.\label{Fig:Jth}]
{ \includegraphics[width=0.5\linewidth]{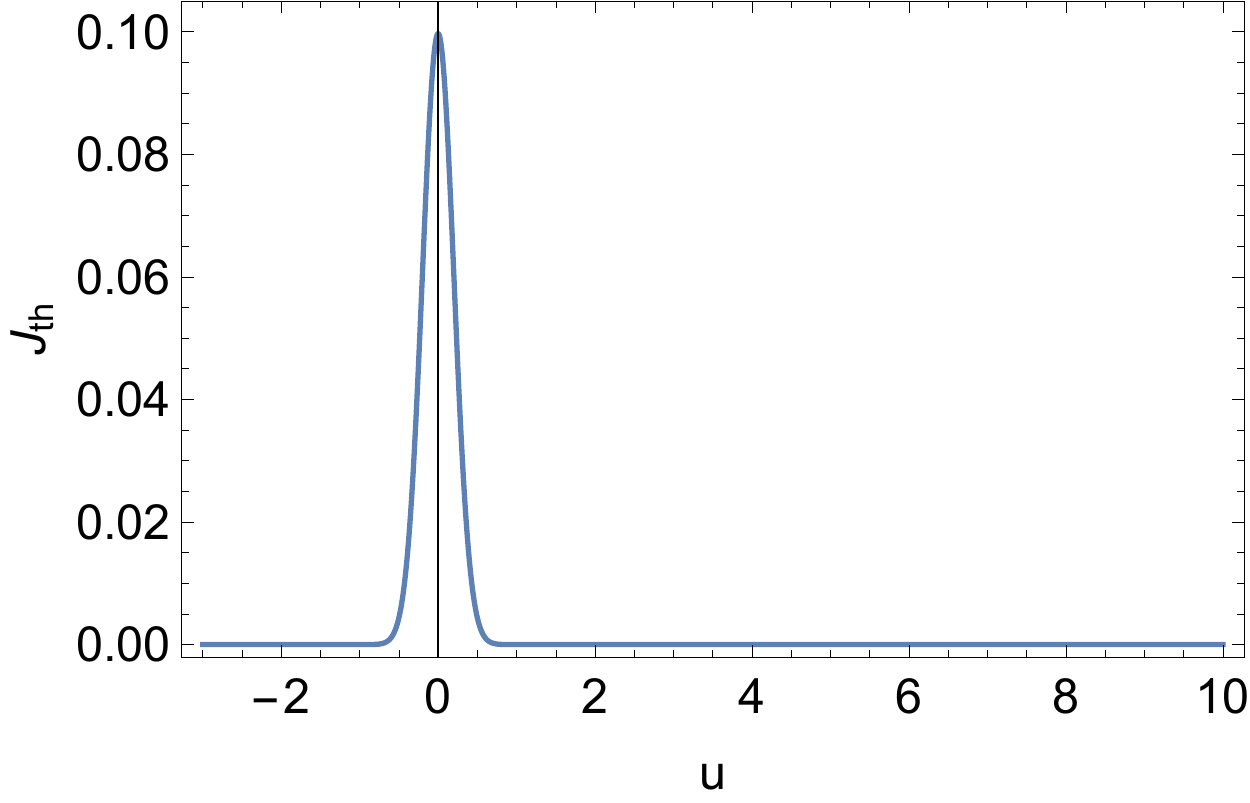}} 
\subfloat[Plot of $O_{th}((\tau(u))$.\label{Fig:Oth}]
{ \includegraphics[width=0.5\linewidth]{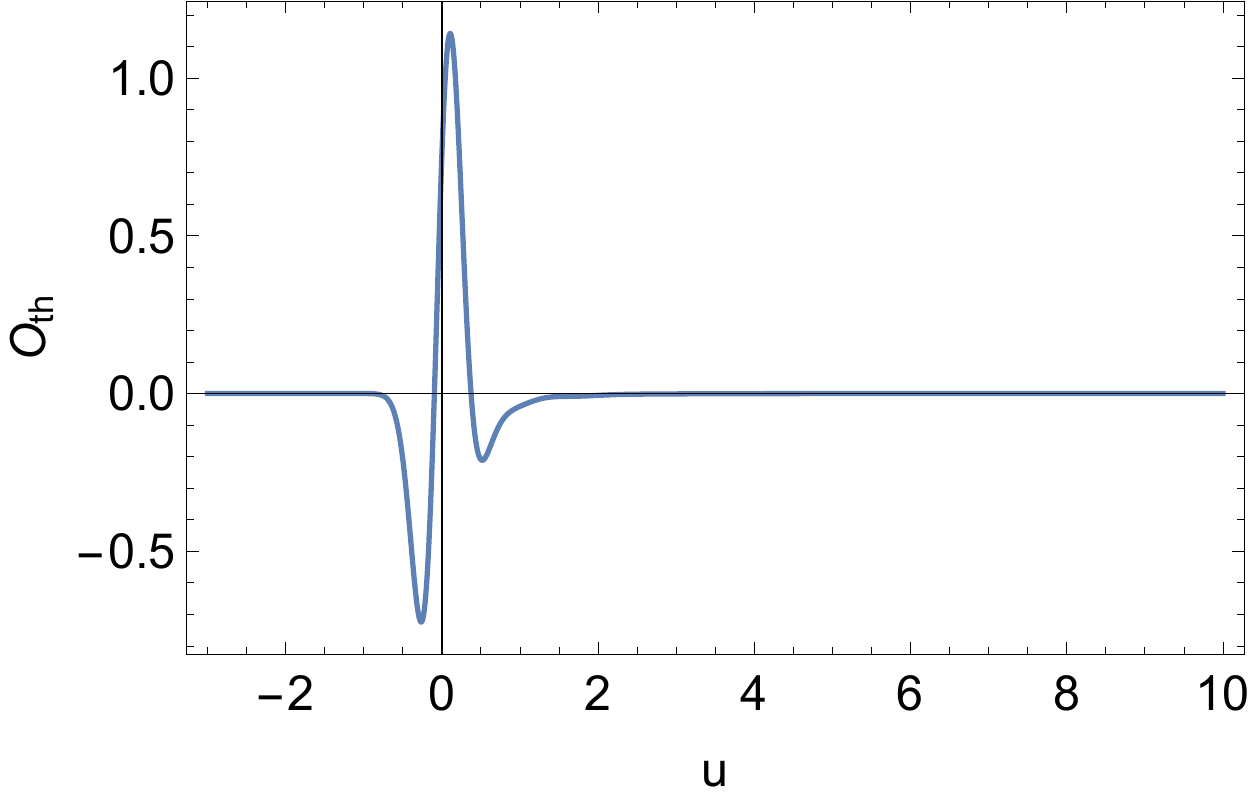}}
\end{center}
\caption{Sources and responses: As expected, the responses die down at late time once the sources vanish.} 
\end{center}
\end{figure}

The chosen Gaussian source $J(u)$ is shown in Fig. \ref{Fig:Ju}. We plot the resulting $O(u)$ in Fig. \ref{Fig:Ou}. After conformal mapping to the state with constant temperature $2\pi$, the source $J_{th}(\tau(u))$ and the response $O_{th}(\tau(u))$ are as shown in Fig. \ref{Fig:Jth} and Fig. \ref{Fig:Oth} respectively. We readily observe that the conformal mapping hardly alters the source and the response.

\begin{figure}
\begin{center}
\begin{center}
\subfloat[Plot of $H_{Sch} = - 1/2 \, M(u)$: This plot is very similar to the case of quenches in higher dimensional holographic systems where $M(u)$ grows but not monotonically. \newline\label{Fig:HSch}]
{\includegraphics[width=\linewidth]{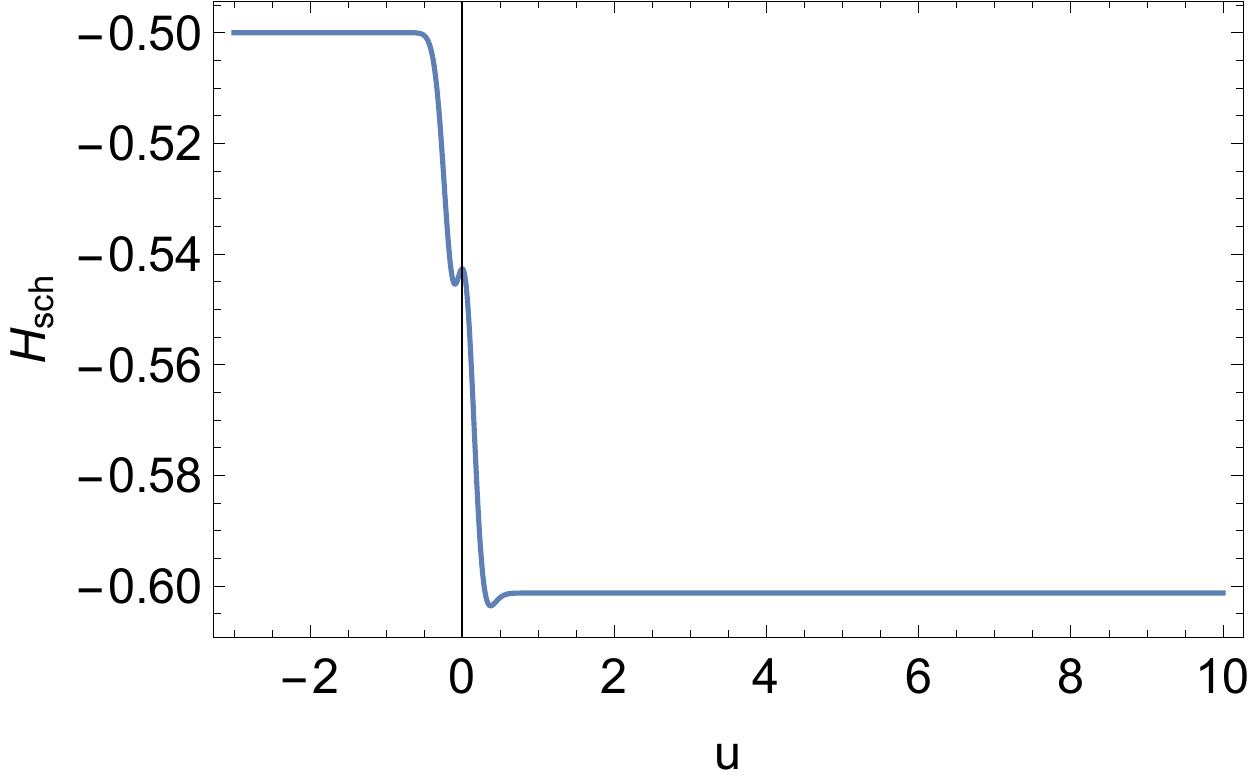}} \quad
\subfloat[Plot of the $SL(2,R)$ charges as a function of time: Note that the final $SL(2,R)$ frame is different since $Q^\pm$ are non-vanishing. \label{Fig:charges}]
{ \includegraphics[width=0.4\textwidth]{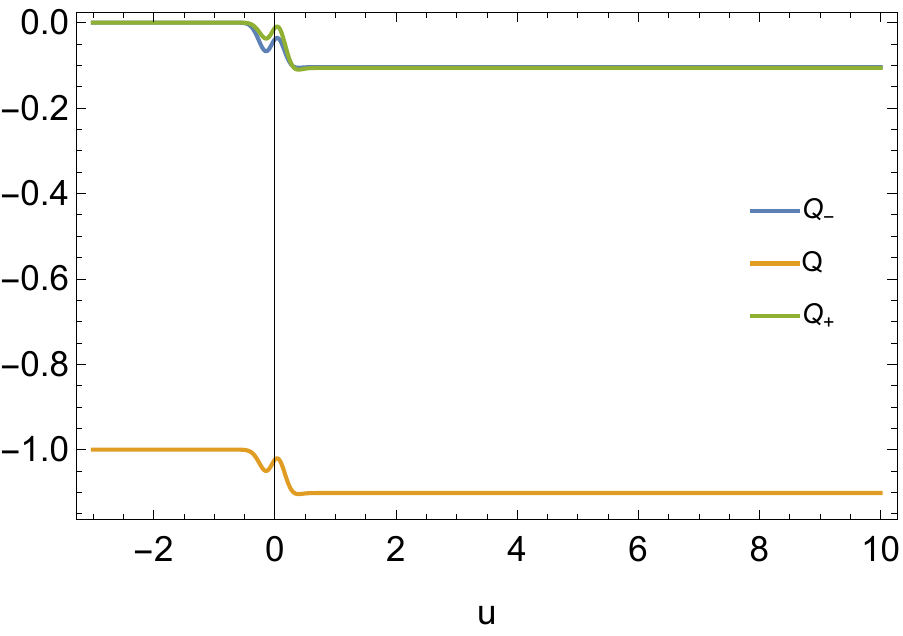}} 
\end{center}
\caption{The time-dependence of the black hole mass and the $SL(2,R)$ charges.} 
\end{center}
\end{figure}

The time-dependence of the black hole mass and the $SL(2,R)$ charges are as shown in Fig. \ref{Fig:HSch} and Fig. \ref{Fig:charges} respectively. Although the black hole mass does not change monotonically just as in the case of higher dimensional analogues, the final black hole mass is significantly bigger than the initial black hole mass. Also the final $SL(2,R)$ frame is different from the initial one.  This $SL(2,R)$ rotation is physically measurable although with some difficulty as it would require measurement of correlation functions $G(u, u')$ with very large $u-u'$ and with $(u+u')/2$ fixed to values when $J$ is large. We thus explicitly find that the quench (pump) leads to formation of soft hair on the black hole represented by $SL(2,R)$ frame rotation.

\begin{figure}
\centering
\includegraphics[width=\linewidth]{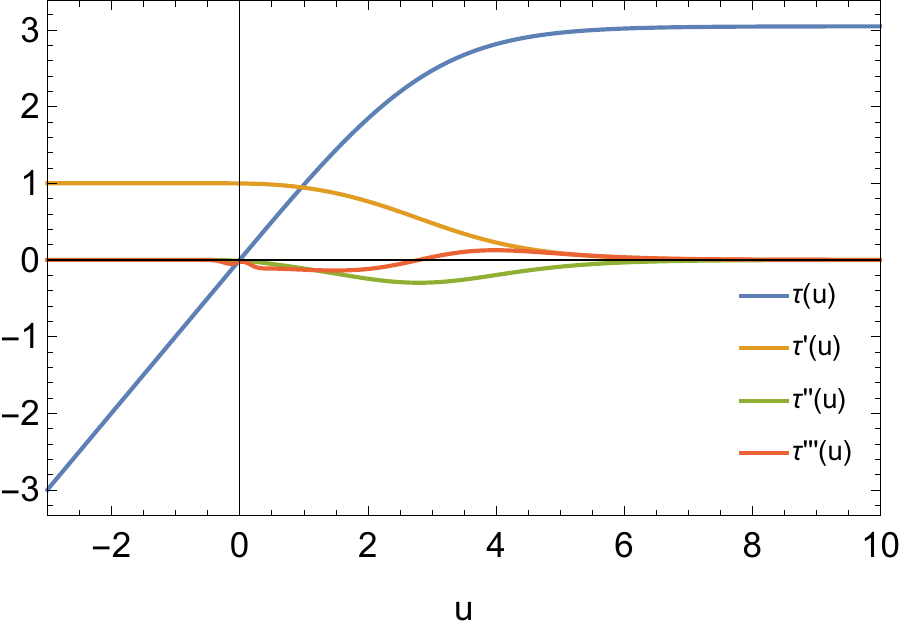}\caption{The plot of $\tau(u)$, which maps the time of the physical state to that of the fixed temperature state, and its derivatives. The generic result is that $\tau (u)$ saturates to a constant and its derivatives vanish.}\label{Fig:holotaus}
\end{figure}

The $SL(2,R)$ charges imply that $\tau'$, $\tau''$ and $\tau'''$ behave as shown in Fig. \ref{Fig:holotaus}. Remarkably, $\tau$ saturates to a constant, so that the map of the time of the physical state to that of the fixed temperature state has a finite endpoint. We observe that $\tau'$ is always positive (ensuring that the map to the time of the fixed temperature state is causal) and $\tau''$ is always negative. 
\paragraph{Verification of the second law:} The quantum quench leads to a transition between two \textit{static} configurations for  \textit{both} the bulk metric and the bulk dilaton. Therefore, the arguments presented in Section \ref{dilpf} imply that the second law should hold, i.e. $\Phi(\lambda)$ should increase monotonically along any smooth null curve with affine parameter $\lambda$. We should choose an appropriate null curve with which we will be able to interpolate between the initial and final thermal entropies monotonically. Such an appropriate choice is the \textit{event} horizon, the smooth null curve (geodesic) interpolating the initial and final horizons at $r_{h\pm} = 1/\sqrt{M_\pm}$ where $M_\pm$ are the final (initial) black hole masses. Evidently from Eq. \eqref{EFMu}, this event horizon can be obtained by solving
\begin{equation}\label{Eq:eventhorizon}
\frac{{\rm d} r}{{\rm d }u} = \frac{1}{2}(M(u) r^2 -1)
\end{equation}
with the boundary condition that $r(u \rightarrow \infty) = 1/\sqrt{M_+}$. Note the event horizon is not determined causally because the final black hole mass $M_+$ is determined by the full history of the quenching protocol.

The dilaton profile $\Phi(r,u)$ during the quench can be readily computed following the algorithm mentioned in the previous subsection. At $u \rightarrow \pm \infty$ it is however easy to see from \eqref{PhisolEFfin} that $\Phi(r, u = \pm \infty) = 3/(2 r)$ because the bulk scalar $\chi$ vanishes. Therefore the value of $\Phi$ on the horizon interpolates between $\Phi_\pm = 3/(2 r_{h\pm})$ where $r_{h\pm} = 1/\sqrt{M_\pm}$  is the location of the horizon at  $u \rightarrow \pm \infty$ when the black hole masses are $M_\pm$.  Furthermore, $M_\pm = \pi^2/\beta_\pm^2 = \pi^2 T_\pm^2$ with $T_\pm$ being the final (initial) temperatures. It follows that $\Phi_\pm \propto (3/2) (\pi T_\pm)$. Identifying the on-shell gravitational (Schwarzian) action with the free energy, we can readily see that the entropy $S_\pm \propto T_\pm$ as should be the case \cite{Maldacena:2016upp}. 

Computing $\Phi(r,u)$ via our numerical algorithm and plotting it on the horizon \eqref{Eq:eventhorizon}, we obtain Fig. \ref{Fig:holoentropy}. It is clear then that the entropy grows monotonically from the initial to the final thermal value. Note that the entropy starts growing much before the quench is significant (around $u = 0$). This peculiarity is due to the non-causal nature of the event horizon on which the dilaton is evaluated.

\begin{figure}
\centering
\includegraphics[width=\linewidth]{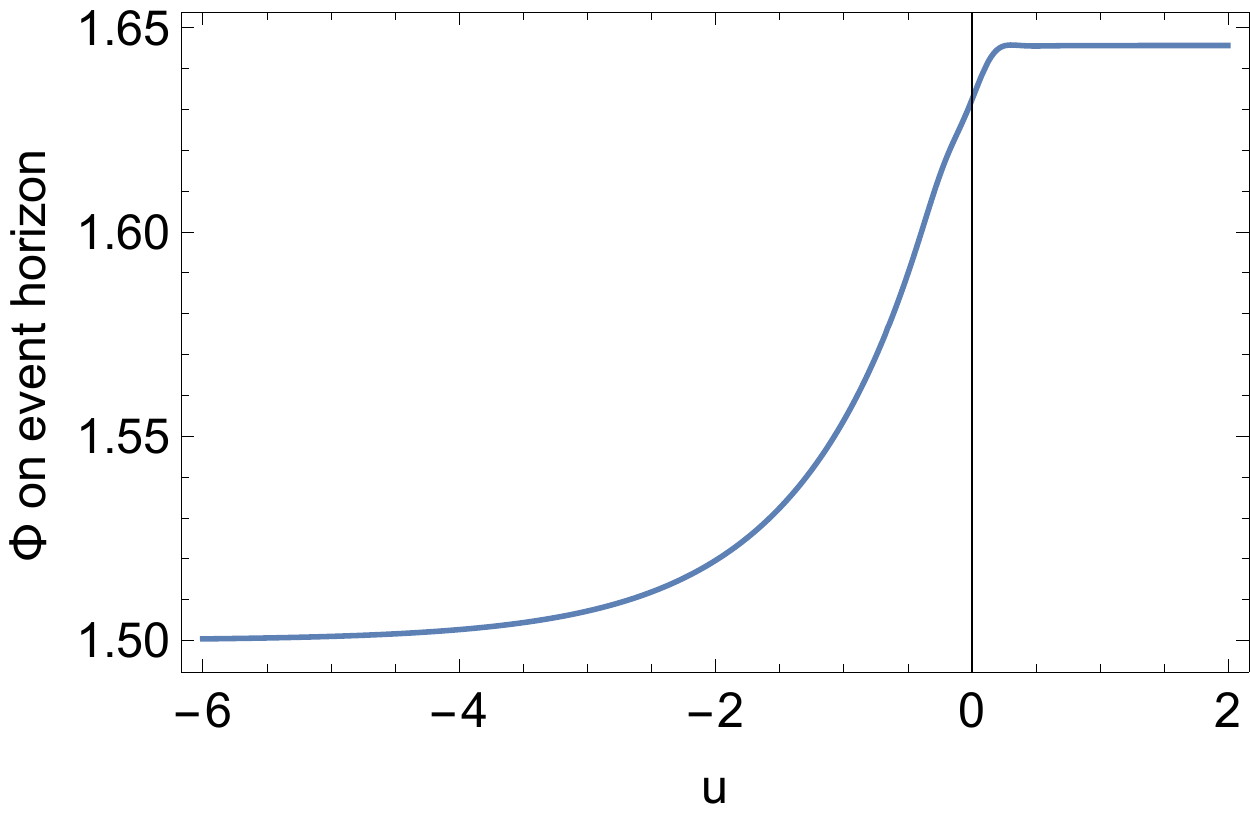}\caption{The dilaton grows on the black hole event horizon monotonically within numerical accuracy interpolating between the thermal limits at early and late times.}\label{Fig:holoentropy}
\end{figure}



Finally, we note that quantum quenches in SYK model have been studied in \cite{Eberlein:2017wah,Bhattacharya:2018fkq}. However, we consider different types of deformations here. It will be interesting to obtain our results using field-theoretic tools.
\section{A semi-holographic model for trapped impurities}
\subsection{Our model}
We will construct a simple semi-holographic model for confined impurities and their mutual strong interactions. The time-dependent position $\vec{X}(u)$ of an impurity can be treated as an extra field in the effective $0+1-$D theory.  When the orbital angular momentum vanishes, the motion is one-dimensional. Here we will restrict ourselves to this simple situation.  

In our model, the strongly interacting $NAdS_2$ holographic sector depicts the dual infrared dynamics of the localized mutual interactions of the impurities confined at the origin $X(u) = 0$. The motion in space of a displaced impurity can be thought of as a deformation of the $NAdS_2$ holographic theory with $X(u)$ representing a self-consistent external source. The center of the force $X(u) = 0$ from the point of view of the $NAdS_2$ holographic sector is then the value of the source for which the deformation to the Schwarzian action vanishes. Since $X(u)$ itself follows Newtonian dynamics, the whole description is semi-holographic \cite{Banerjee:2017ozx,Kurkela:2018dku}, i.e. holography with a self-consistent dynamical source at the boundary and with a total conserved energy.

The effective string tension of the confining force is thus the self-consistent expectation value of an operator $O$ in the $NAdS_2$ holographic theory. The confining potential therefore takes the form
\be\label{V}
V = \lambda O(u) X(u)
\ee
where $\lambda$ is a dimensionful hard-soft coupling constant. Then $\lambda X(u)$ should be identified with the source $J(u)$ (non-normalizable mode) of the bulk scalar field $\chi$ dual to the operator $O(u)$. Requiring that the holographic theory suffers only an irrelevant deformation about the Schwarzian action and that it retains $SL(2,R)$ invariance in the large $N$ limit (classical gravity approximation) imply that $O(u)$ must have scaling dimension $\Delta$ such that $1 < \Delta < 3/2$. The mass of the dual bulk field $\chi$ should satisfy $0 < m^2 l^2 < 3/4$ since $m^2 l^2  = \Delta(\Delta -1)$ with $l$ being the radius of $AdS_2$ (we set $l=1$), and its asymptotic expansion should be
\be
 \chi(r, u) \approx \lambda X(u) r^{1-\Delta} + \cdots.
 \ee

We will now construct the full self-consistent dynamics such that a total conserved energy exists. Let's start with the boundary field $X(u)$.  Newton's law corresponding to the potential \eqref{V} readily gives
 \be\label{Newton}
 m_i X''(u) = - \lambda O(u)
 \ee
where $m_i$ is the mass of the impurity.  Then the \textit{kinetic energy} is
\be\label{Tu}
H_{kin} = \frac{1}{2}m_i X'(u)^2,
\ee
 which satisfies 
 \be\label{T}
H_{kin}' = -  \lambda O(u) X'(u).
 \ee

The algorithm for determining $O(u)$ along with the time reparametrization $\tau(u)$  (equivalently the mass $M(u)$ of the $AdS_2$ black hole) has been discussed before.  Assembling our previous results, we quote the equation of motion \eqref{WIH} for $\tau(u)$
 \begin{align}\label{HolWI}
 \Big( Sch(\tau(u),u) -\frac{1}{2}\tau'(u)^2 &- \lambda(\Delta -1)  X(u) O(u)\Big)' \nonumber\\
 &= \lambda O(u) X'(u).
 \end{align}
 Above $O(u)$ should be obtained self-consistently by solving the Klein-Gordon equation in the $AdS_2$ black hole background \eqref{EFM1} with $M(u) =1$ but with source specified by
 \be
 X_{th}(\tau(u)) = X(u)\tau'(u)^{\Delta -1}.
 \ee
The response to this source is $O_{th}(\tau(u))$ from which we can extract $O(u)$ utilizing the relation
 \be\label{O(u)}
 O(u) = O_{th}(\tau(u)) \tau'(u)^\Delta. 
 \ee
 The equations \eqref{Newton} and \eqref{HolWI} thus completely specify the semi-holographic dynamics.
 
 We readily note that \eqref{T} and \eqref{HolWI} imply the existence of a total conserved energy $H_{tot}$ satisfying
 \be
 H_{tot}' = 0, 
 \ee
 and which is explicitly given by
 \begin{align}\label{Htot}
 H_{tot} &= H_{kin} + Sch(\tau(u),u) -\frac{1}{2}\tau'(u)^2 \nonumber\\
 &- \lambda (\Delta -1) X(u) O(u)\nonumber\\
 &= H_{kin} - \frac{1}{2} M(u) -  \lambda (\Delta -1) X(u) O(u).
 \end{align}
 In the second line, we have used  \eqref{SchMu} relating $M(u)$ and $\tau(u)$.\footnote{Note in order to have the right dimensions $M(u)$ should really be $M(u) c_{IR}^2$ in \eqref{Htot} where $c_{IR}$ is the effective velocity for causal propagation in the infrared sector and is not necessarily the speed of light. We use the natural units $c_{IR}=1$ here.} We readily see that the terms other than $H_{kin}$ can be interpreted as a self-consistent effective potential:
 \be
 V_{eff} = - \frac{1}{2} M(u) - \lambda (\Delta -1) X(u) O(u).
 \ee
 \Ayan{The action of the full system from which all equations of motion i.e. \eqref{Newton} and \eqref{HolWI} follow is thus}
 \begin{equation}\label{semiholo-action}
 S = \frac{1}{16 \pi G}\int {\rm d}u \frac{1}{2}m_i X'^2 - S_{\rm on-shell}^{\rm grav}[J(u) = \lambda X(u)] 
 \end{equation}
 \Ayan{where $S^{grav}$ is given by \eqref{onshell-full}. This action should be viewed as a functional of $X(u)$ and $t(u)$. Noting that}
 \begin{equation*}
16 \pi G \frac{\delta S_{grav}}{\delta X(u)} = 16 \pi G \frac{\delta S_{\rm on-shell}^{\rm grav}}{\delta J(u)}\frac{\delta J(u)}{\delta X(u)} = \lambda O(u)
 \end{equation*}
 \Ayan{we readily find that extremizing \eqref{semiholo-action} w.r.t. $X(u)$ yields the Newtonian equation \eqref{Newton}. On the other hand, extremizing $S_{\rm on-shell}^{\rm grav}$ w.r.t. $t(u)$ yields \eqref{HolWI} as we have noted before.}
 
 \Ayan{The relative sign between the two terms in the full action given by Eq. \eqref{semiholo-action} can look strange. However, as explained in \cite{Brown:2018bms}, this relative sign appears in the context of effective JT gravities when we trade in kinetic energy of an extraneous degree of freedom for an effective potential energy (see Appendix A of \cite{Brown:2018bms} for a cogent explanation using the analogy of the classic central force problem). In our case, the full action can be regarded as the action on the worldline of the displaced impurity with the gravitational $NAdS_2$ system providing a self-consistent effective potential energy. Our full action has a higher dimensional origin like the examples discussed in \cite{Brown:2018bms} but note that it cannot be embedded in a higher dimensional holographic setup.}
 
 The equilibrium solution for the above problem is $X(u) = 0$ where the confining force vanishes\footnote{If $X(u) \neq 0$, it will generate $O(u)$ via the Klein Gordon equation for the dual bulk scalar field.} and in which the bulk is thermal at the ambient medium temperature so that $M(u)$ is a constant. The bulk scalar vanishes as does $O(u)$.  Our initial conditions are set by such an equilibrium configuration. To usher in time-dependence, we consider an impulse generated by an external force $F(u)$ which originates from a fluctuation in the medium where the impurities are living and which is of the form of a delta function,\footnote{It has been shown in \cite{Banerjee:2016ray} that under similar circumstances the delta function limit where the width of a narrow Gaussian vanishes keeping the impulse fixed can be taken smoothly in numerical holography.} i.e.
 \be
 F(u) = m_i v_0 \delta(u - u_0).
 \ee
  The equation for $X(u)$ given by \eqref{Newton} should then be replaced by
 \be\label{Newton-new}
 m_i X''(u) = F(u) -\lambda O(u).
 \ee
 The full system exists in the equilibrium configuration for $u < u_0$. At $u= u_0$, the impulse generated by $F(u)$ will impart a finite velocity $X'(u_0) = v_0$ thus infusing energy into the system. The total energy $H_{tot}$ given by \eqref{Htot} will be conserved for $u> u_0$ when $F(u)$ vanishes. Setting the initial temperature to $\beta^{-1} = 1/(2\pi)$ as before by utilizing scaling symmetry and $m_i =1$ for convenience, the time-evolution will be determined by the parameters $v_0$ and the hard-soft coupling $\lambda$.
 
 We can solve for $\tau(u)$ and $O(u)$ following our algorithm as detailed in Section \ref{algo}. The only difference is that unlike before the source $X(u)$ will not be a predetermined function but should be co-evolved according to \eqref{Newton-new}. \Ayan{As noted before, we are free to choose our initial $SL(2,R)$ frame because an overall time-independent $SL(2,R)$ transformation of the full solution of $t(u)$ (and therefore $\tau(u)$) has no physical effect on the observables. Therefore, we choose the initial $SL(2,R)$ charges and also $\tau(u)$ according to \eqref{Q-init}. Additionally, in the detailed seven step algorithm described in Section \ref{algo}, we simply add a new step between the sixth and seventh: we update $X(u)$ using the known $X'(u)$ at the previous instant and update $X'(u)$ according to \eqref{Newton-new} using $O(u)$ at the present instant at each time step.} For concreteness, we will set the scaling dimension $\Delta$ of $O$ to be $5/4$.  We will also assume that $v_0 > 0$ because we want to investigate how far the impurity can be pushed from the center of the confining force.
 
Furthermore, it is not difficult to see that the sign of $\lambda$ is not relevant in our model. Note that the action of the bulk scalar field is quadratic. Since the source of the bulk scalar is $J(u) = \lambda X(u)$, it follows that the response $O(u)$ will be odd in $\lambda$.  Furthermore the interaction term $\lambda X(u) O(u)$,  the confining force $\lambda O(u)$ in \eqref{Newton}, etc. are then even in $\lambda$. We can therefore choose $\lambda > 0$ without loss of generality.
 
 \subsection{Non-equilibrium phase transitions}
 We explore\footnote{We thank Alexandre Serantes for several insightful comments which have improved the presentation and also our understanding of the results of this subsection.} the semi-holographic model described above numerically by varying the initial velocity $v_0$ and the hard-soft coupling $\lambda$. For the following discussion, we will split the total conserved energy $H_{tot}$ in \eqref{Htot} into (i) the kinetic energy of the particle $H_{kin}$ as defined in \eqref{Tu}, (ii) the black hole mass term $H_{sch} = - 1/2\,  M(u)$ and (iii) the hard-soft interaction energy $H_{int} = -\lambda (\Delta -1) X O = - \lambda/4 \,  X O$. 
 
 As far as we have investigated, we find that for any value of $v_0$ and $\lambda$, the mass of the hole $M(u)$ increases (i.e. $H_{sch}$ decreases), and the interaction energy $H_{int}$ is positive \textit{at early time}. It then follows from total energy conservation that the particle kinetic energy $H_{kin}$ increases initially, i.e. the particle undergoes acceleration. Remarkably, the mass of the black hole $M(u)$ \textit{always} goes to zero at very late time and the total energy is fully transferred either to the particle kinetic energy $H_{kin}$ or to the interaction energy $H_{int}$ which reduces to a self-consistent confining potential energy. 
 
 Furthermore, the particle \textit{always} decelerates at late time. When the final total energy transfer goes to its kinetic energy, it reaches a terminal velocity $v_f$ which is less than its initial velocity $v_0$. Energy conservation implies that 
 \be
 \frac{1}{2} m_i v_f^2 = \frac{1}{2}  m_i v_i^2 - \frac{1}{2} M_o,
 \ee 
 where $m_i$ is the mass of the particle (impurity) and $M_0$ is the initial black hole mass. The above relation simply equates the initial and final total energies and determines $v_f$. (Note that the initial interaction energy is zero because the particle starts from the center $X = 0$ initially.) When the final transfer of the total energy  goes to the interaction energy $H_{int}$, the particle comes to a full stop as its kinetic energy vanishes. 
 
Which of these two final outcomes is realized simply follows from the observation that after sufficiently long time, the interaction energy $H_{int}$ is \textit{always} negative. Therefore, it either goes to zero from below when the final outcome is that the total energy is transferred to the particle kinetic energy or saturates to a negative constant if the final outcome is otherwise. The first outcome is possible if and only if the total energy is positive since the kinetic energy is always positive. In the other case, the total  energy has to be negative. Since the initial interaction energy is zero as noted above, the total (conserved) energy $H_{tot}$ is simply given by 
 \be
 H_{tot} = \frac{1}{2} m_i v_0^2 - \frac{1}{2} M_0
 \ee
 as the sum of initial values of the kinetic energy and $H_{sch}$. The final outcome of transfer of total energy to the kinetic energy $H_{kin}$ then happens when $H_{tot} >0$ i.e. for\footnote{Note that both sectors can have different fundamental speeds of causal propagation, with $c_{IR} < c_{UV}$. The non-relativistic limit for the boundary dynamics apply when the particle speed is much less than $c_{UV}$. Note that below and elsewhere $\sqrt{M_0/m_i}$ should actually be $\sqrt{M_0/m_i}c_{IR}$. If $c_{IR} = c_{UV}$, then our model applies only if $M_0 \ll m_i$.} 
 \be
 v_0 > \sqrt{\frac{M_0}{m_i}}.
 \ee
 The other final outcome of transfer of total energy to the potential energy $H_{int}$ occurs when
 \be
 v_0 < \sqrt{\frac{M_0}{m_i}}.
 \ee
When $v_0 = \sqrt{M_0/m_i}$ (i.e. $H_{tot} = 0$), both $H_{int}$ and $H_{kin}$ vanish at late time along with $H_{sch}$. We will say more about this special case later.
 
 A closer look at the bulk solution reveals a more interesting phase transition which depends on whether the mass of the black hole $M(u)$ always stays positive throughout the time evolution, or whether it undergoes one or more oscillations before it finally goes to zero. The first case occurs for
 \be
 v_0 > v_c(\lambda) > \sqrt{\frac{M_0}{m_i}}
 \ee
 implying that in this phase the total energy always goes to the kinetic energy of the particle which therefore never stops. For $v_0 < v_c(\lambda)$, the final outcome can then be either attainment of a terminal velocity or full stopping depending on whether $v_0 > \sqrt{M_0/m_i}$ or otherwise. {The crucial point is that for $v_0 > v_c(\lambda)$, the mass of the black hole always remains positive and eventually vanishes at late time while for $v_0 < v_c(\lambda)$ the mass of the black hole becomes zero in \textit{finite} time.  In the latter case, the black hole mass then becomes negative and the magnitude diminishes monotonically at late time, or the mass then oscillates about zero at least once before vanishing at late time.}
 
 It is somewhat surprising that although in the pure holographic case the final mass of the black hole is greater than its initial mass as we have reported before, in the semi-holographic case the final mass at very long time is always zero. A similar phenomenon of disappearing horizon has been observed before in \cite{Kourkoulou:2017zaj}.\footnote{The interpretation of this result was attributed to work being done by the black hole rather than on it, see also \cite{Dhar:2018pii}. In our case, a similar interpretation is naturally obtained via the virtue of total energy conservation.} In  semi-holography, the late time behavior is not controlled by the quasi-normal modes of the individual systems since the actual collective modes are hybrid excitations of both systems (see \cite{Kurkela:2018dku} for a detailed exposition of collective modes in the case of a two fluid system).\footnote{Also note that quasi-normal mode in holography results from imposing both the Dirichlet boundary condition at asymptotia and the infalling boundary condition at the horizon. In semi-holography, the Dirichlet boundary condition is not imposed. Instead it is specified by the dynamics of the self-consistent source. Therefore, the late time behavior should not be determined by the usual quasi-normal mode. In fact, if the quasi-normal mode governed the late-time behavior, then $O(u)$ could not have vanished when $X(u)$ grows linearly at late time in the case of the first phase. Then $H_{int}$ also could not have decayed  at late time in the first phase as observed.} This is why the long term behavior of a semi-holographic system can be very different from that of a purely holographic system. 
 
 However in higher dimensions, a similar simulation shows that if the boundary fields do not have many degrees of freedom and only scalar hard-soft couplings are present, the black hole sucks up all the energy depleting the boundary sources \cite{Ecker:2018ucc}. The case of JT gravity is peculiar and we also note that it cannot be embedded in a higher dimensional setup as the dilaton does not couple to matter directly.  
  
 \subsubsection{An illustrative example of phase one behavior}\label{sec:phase1}
 Here we will study the case of $v_0 = 2.0$ for $\lambda = 0.4$ as an example of phase one behavior.  In this example, the mass of the black hole is positive definite and it vanishes at long time. So $H_{sch}$ is negative definite and it goes to zero from below. Plots of $H_{kin}$, $H_{sch}(u)$ and $H_{int}(u)$ are shown in Fig. \ref{Fig:EnergiesPhase1}. Indeed one observes that $H_{sch}$ and $H_{int}$ both vanish at long time while $H_{kin}$ stabilizes conserving total energy. It is instructive to study the time-dependence of the (gravitational) $SL(2,R)$ charges as shown in Fig. \ref{Fig:ChargesPhase1} -- all of them diverge at long time although the Casimir, which is proportional to the black hole mass, goes to zero. {We show below that the late-time behavior of the $SL(2,R)$ charges is captured via an $SL(2,R)$ invariant exponent from which we can recover the knowledge of the initial conditions for the particle at the bourndary.}
 
We also plot $X(u)$ and $X_{th}(\tau(u))$ in Fig. \ref{Fig:XuXthPhase1}, $O(u)$ in Fig. \ref{Fig:OuPhase1} and $O_{th}(\tau(u))$ in Fig. \ref{Fig:OthPhase1}. We find that $X(u)$ attains a terminal velocity i.e. grows linearly at late time although remarkably the physical response $O(u)$ determining the string tension in the confining force vanishes fast enough so that the product $X(u) O(u)$ and hence $H_{int}$ also vanishes. In contrast, neither $X_{th}(\tau(u))$ nor $O_{th}(\tau(u))$ decays at late time, but $H_{int}$ is proportional to $ \tau'  X_{th}O_{th}$ and in this \textit{picture} its decay is ensured by the behavior of $\tau'$. Also note that $O(u)$ stays positive after some initial time so that the force on the impurity (see Eq. \eqref{Newton}) is indeed confining in the long run and the interaction energy $H_{int}$ goes to zero from below as previously claimed.

\begin{figure}
\begin{center}
\begin{center}
\subfloat[Plot of energies as function of time: Note that the total energy $H_{tot} = H_{kin} + H_{int} + H_{sch}$ is conserved after the initial impulse and is finally transferred to $H_{kin}$, the particle kinetic energy. The mass of the black hole $M = - 2 H_{sch}$ remains positive and decays to zero eventually. \label{Fig:EnergiesPhase1}]
{ \includegraphics[width=\linewidth]{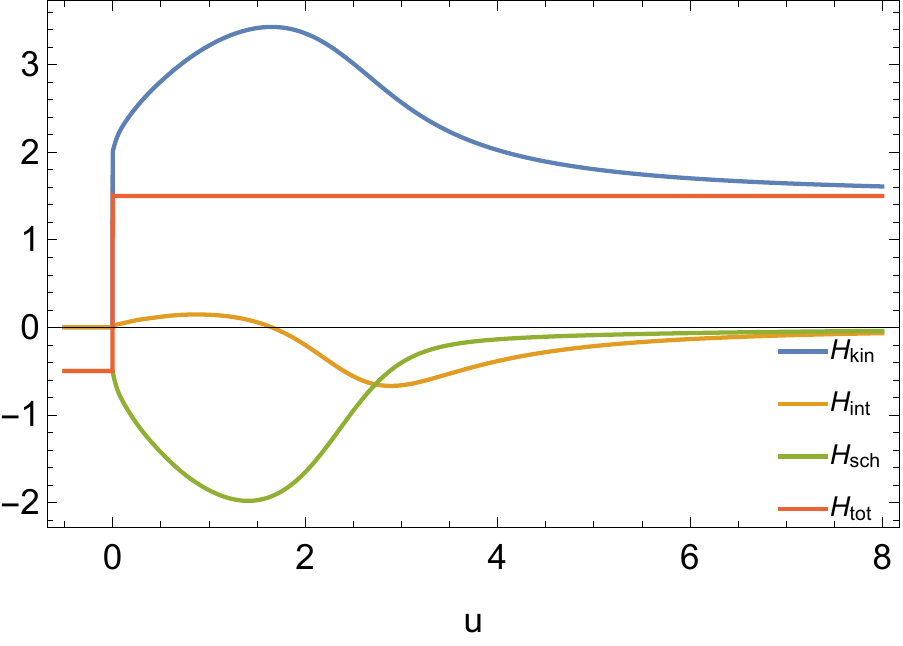}} \;
\subfloat[Plot of the $SL(2,R)$ charges as a function of time in the first phase. Although all of them diverge at late time, their Casimir (and thus the black hole mass) vanishes.\label{Fig:ChargesPhase1}]
{  \includegraphics[width=\linewidth]{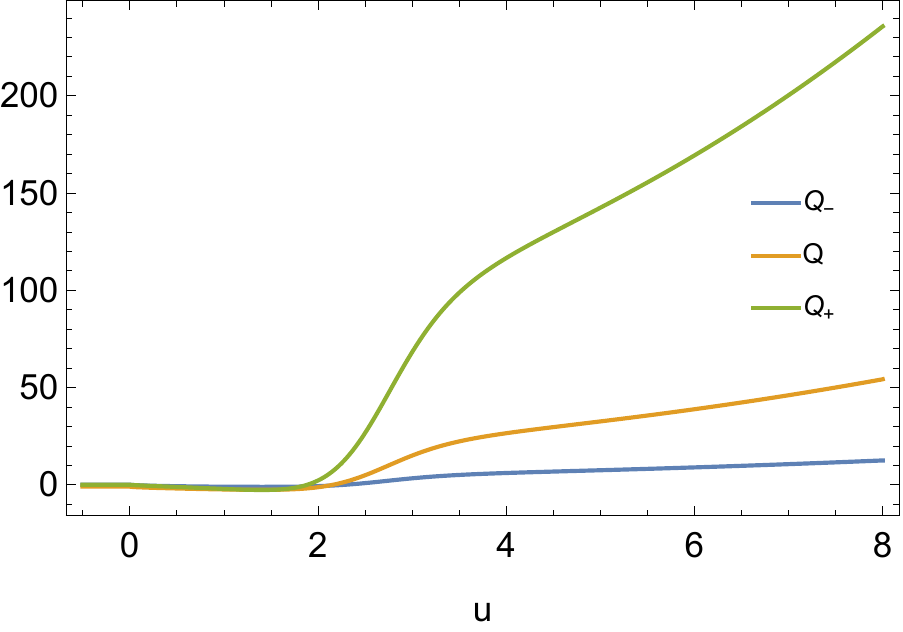}} 
\end{center}
\caption{The plots for energies and $SL(2,R)$ charges for $v_0=2.0$ and $\lambda=0.4$} 
\end{center}
\end{figure}

\begin{figure}
\begin{center}
\begin{center}
\subfloat[Plot of $X(u)$ and $X_{th}(\tau(u))$ as functions of time. Note $X(u)$ eventually reaches linear growth regime implying that the particle reaches a terminal velocity. $X_{th}(\tau(u))$, the source conformally mapped to a black hole of unit mass, saturates to a constant. \label{Fig:XuXthPhase1}]
{ \includegraphics[width=\linewidth]{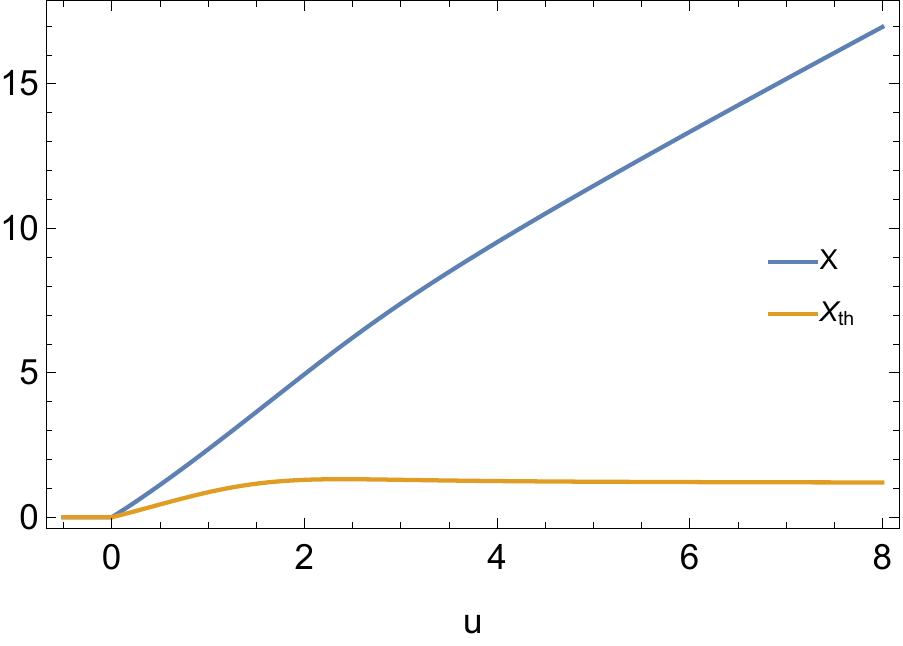}}
\\
\subfloat[Plot of $O(u)$ as a function of time. The eventual rapid decay of $O(u)$ ensures that $H_{int} \propto X(u) O(u)$ vanishes at long time.\label{Fig:OuPhase1}]
{ \includegraphics[width=\linewidth]{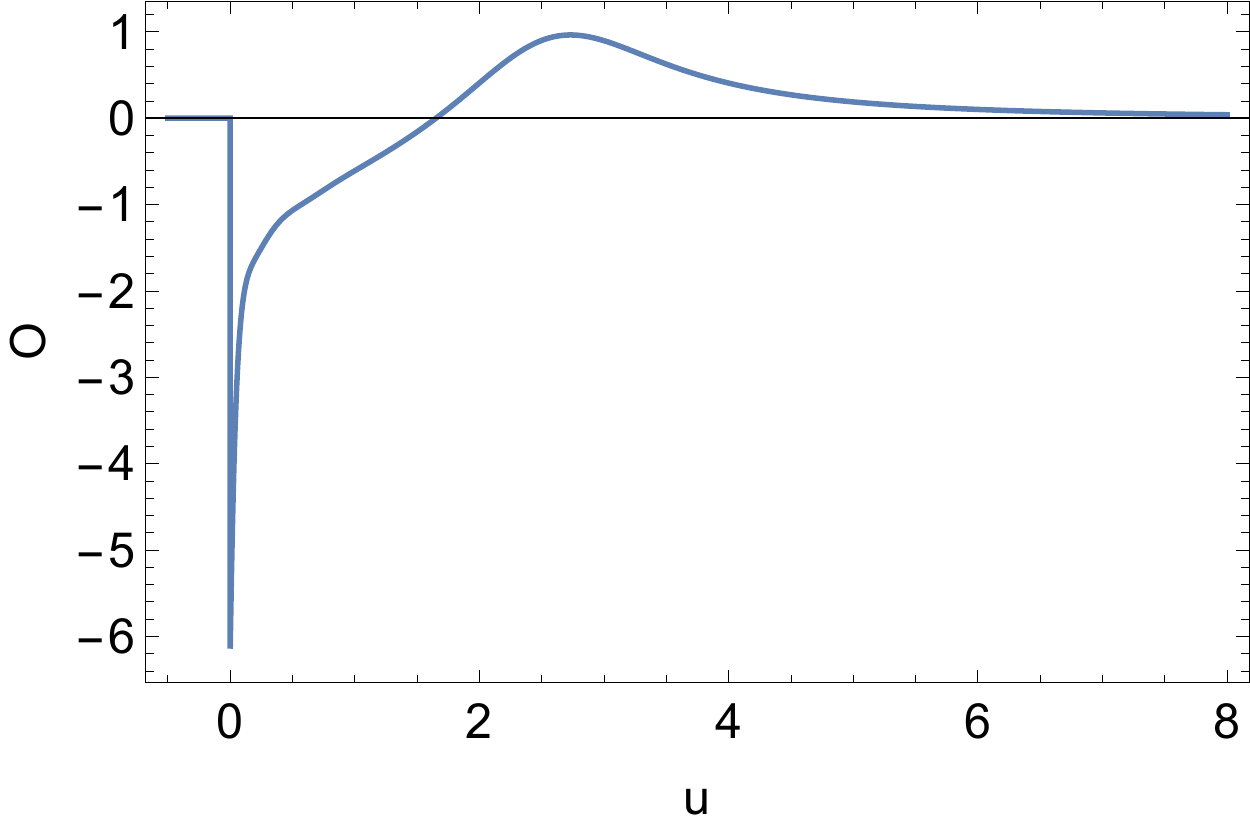}}
\\
\subfloat[Plot of $O_{th}(\tau(u))$: Note as $X_{th}(\tau(u))$ saturates, $O_{th}(\tau(u))$ grows with time. However, the rapid decay of $\tau'$ ensures that $H_{int}\propto \tau'(u)X_{th}(\tau(u)) O_{th}(\tau(u))$ also decays in this frame.\label{Fig:OthPhase1}]
{ \includegraphics[width=\linewidth]{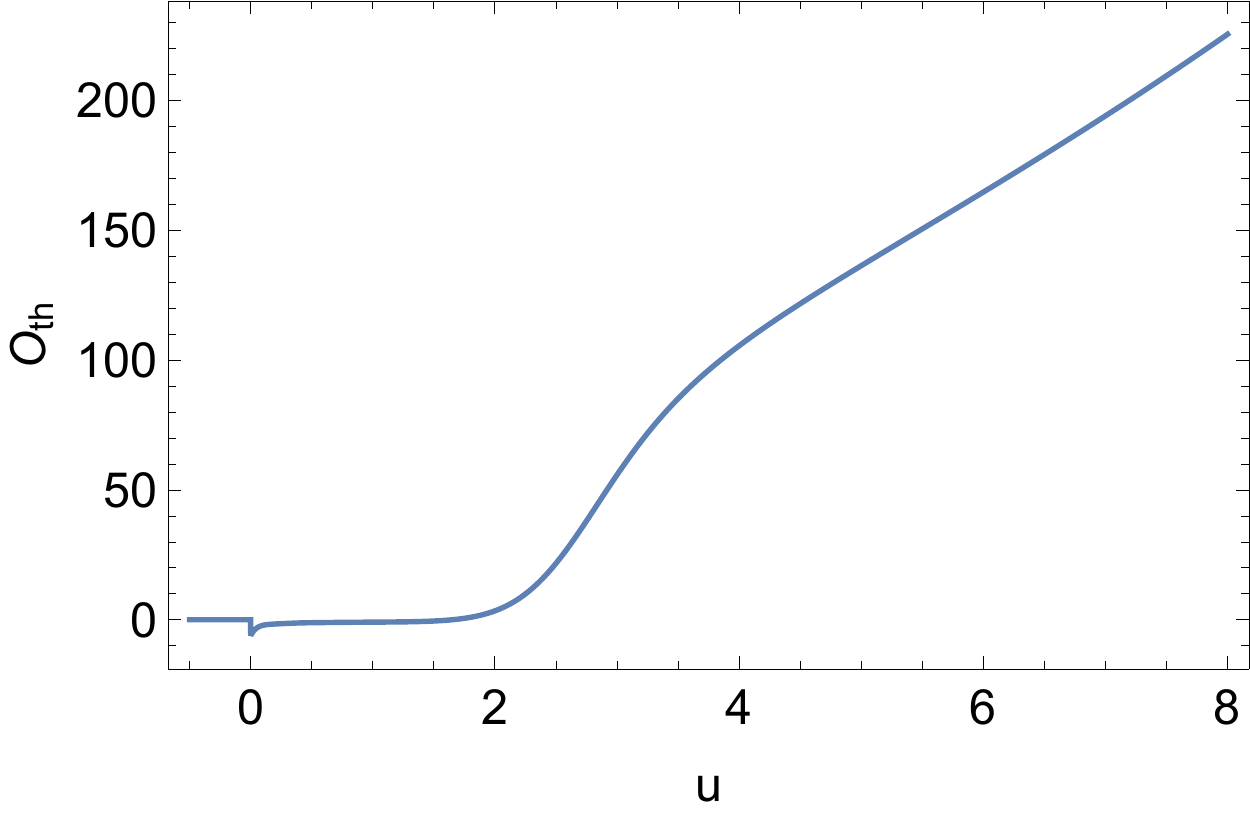}} 
\end{center}
\caption{The plots of sources and responses for $v_0=2$ and $\lambda=0.4$.} 
\end{center}
\end{figure}

The bulk metric which corresponds to the observer's time takes  the form \eqref{EFMu} in the ingoing Eddington-Finkelstein coordinates $r$ and $u$ with $M(u) = - 2 H_{sch}(u)$. The profile of the dilaton $\Phi(r,u)$ can be readily obtained following the method of Section \ref{dilpf} (see also the discussion on the alternative numerical algorithm in Section \ref{algo}). {The dilaton remains finite in the entire physical patch covered by the $(r,u)$ coordinates as far as we have studied. \footnote{In order to understand the behavior at $r = \infty$, it is instructive to change to $\rho, \tau$ coordinates where it gets mapped to $\rho = - \tau'^2/\tau''$ according to \eqref{EFdiffeo}. Referring to Fig. \ref{Fig:taus} we note that $- \tau'^2/\tau'' > 0$ always.} It turns out that the dilaton vanishes at a locus in the interior. One may think that the locus where the dilaton vanishes could also be a singularity since the effective Newton's constant diverges there. We do not think this to be the case in our model of JT gravity. Since the dilaton does not couple to matter and gravity has no bulk propagating mode, the vanishing of $\Phi$ does not lead to any singular propagator. There are of course physical fluctuations of the Schwarzian part of the action (leading to so-called \textit{boundary gravitons}) but its pre-factor is the constant $\overline{\phi}_r$ and it is well behaved even if $\Phi$ vanishes in the interior. Furthermore, the bulk mutter fluctuations couple to these boundary gravitons only. \footnote{Note the same could also be said if the dilaton diverges at a locus in the bulk because an infalling particle which does not couple to the dilaton will not see this singularity. Pathologies, if any, will only be visible by the time-reparametrization function. As far as we are aware, all our solutions are non-pathological and also such singularities are absent in the physical patch.}}



{It is interesting to ask how the physical patch covered by the $r,u$ coordinates fit in the Poincar\'{e} patch covered by the Fefferman-Graham coordinates. We readily find that} $r = \infty$ is within the Poincar\'{e} patch. Referring to Eqs. \eqref{EFdiffeo} and \eqref{staticdiffeo} we find that $r = \infty$ maps to
\begin{align}
z &= G(u),\nonumber\\  G(u) &= \frac{1}{2}\Bigg[\tanh\left(\frac{\tau(u)}{2} + {\rm arctanh} \left(- \frac{\tau'(u)^2}{\tau''(u)}\right)\right)\nonumber\\
&\qquad\qquad\qquad\qquad-\tanh\left(\frac{\tau(u)}{2}\right)\Bigg].
\end{align}
\normalsize
As evident from Fig. \ref{Fig:taus}, $\tau(u)$ saturates to a constant at large time while $\tau'$, $\tau''$ and $\tau'''$ decay to zero quite similarly to the case of the pure holographic quench. Also, $\tau'$ is always positive (otherwise the map to the time of the fixed temperature state would not have been causal), $\tau''$ is always negative, and finally $\tau'^2/\tau''$ is always negative and decays to zero at large time as well (see Fig. \ref{Fig:tp2tpp}). As a result, $G(u)$ is always finite and vanishes at very long time  (see Fig. \ref{Fig:Gu}). {Therefore, $r= \infty$ eventually reaches the boundary.}



\begin{figure}
\begin{center}
\begin{center}
\subfloat[Plots of $\tau$, $\tau'$, $\tau''$ and $\tau'''$ vs $u$. Due to our choice of initial $SL(2,R)$ frame, $\tau' = 1$ (so $\tau$ is linear in $u$) for $u< 0$ (before the kick), while $\tau'' = \tau''' = 0$. Note $\tau$ saturates at late time, while all its derivatives vanish. \label{Fig:taus}]
{ \includegraphics[width=\linewidth]{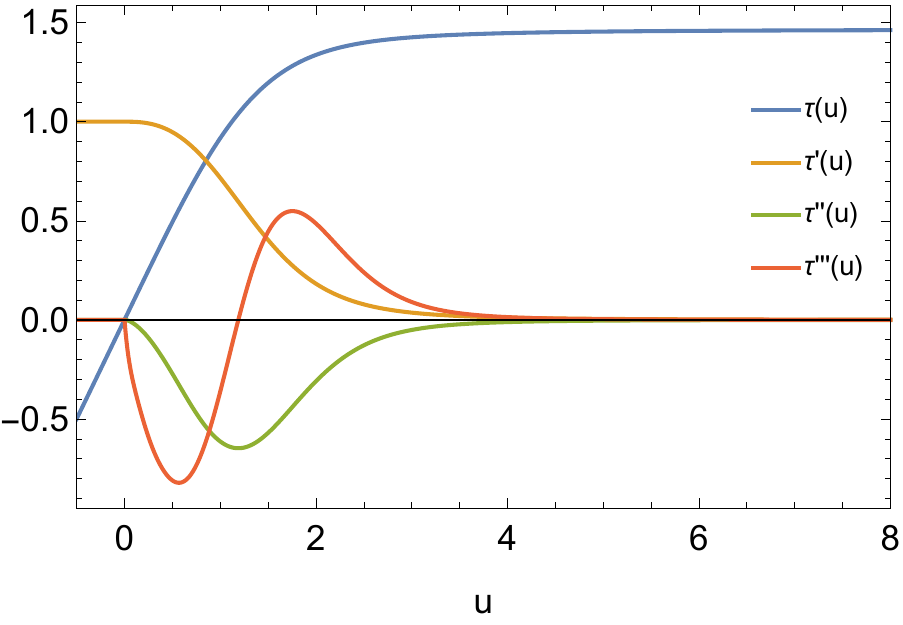}}  \;\;
\subfloat[$\tau'^2/\tau''$ as a function of time. Clearly $\tau'^2$ decays faster than $\tau''$ with time. Also $\tau'^2/\tau''$ is always negative since $\tau''$ is so.\label{Fig:tp2tpp}]
{ \includegraphics[width=\linewidth]{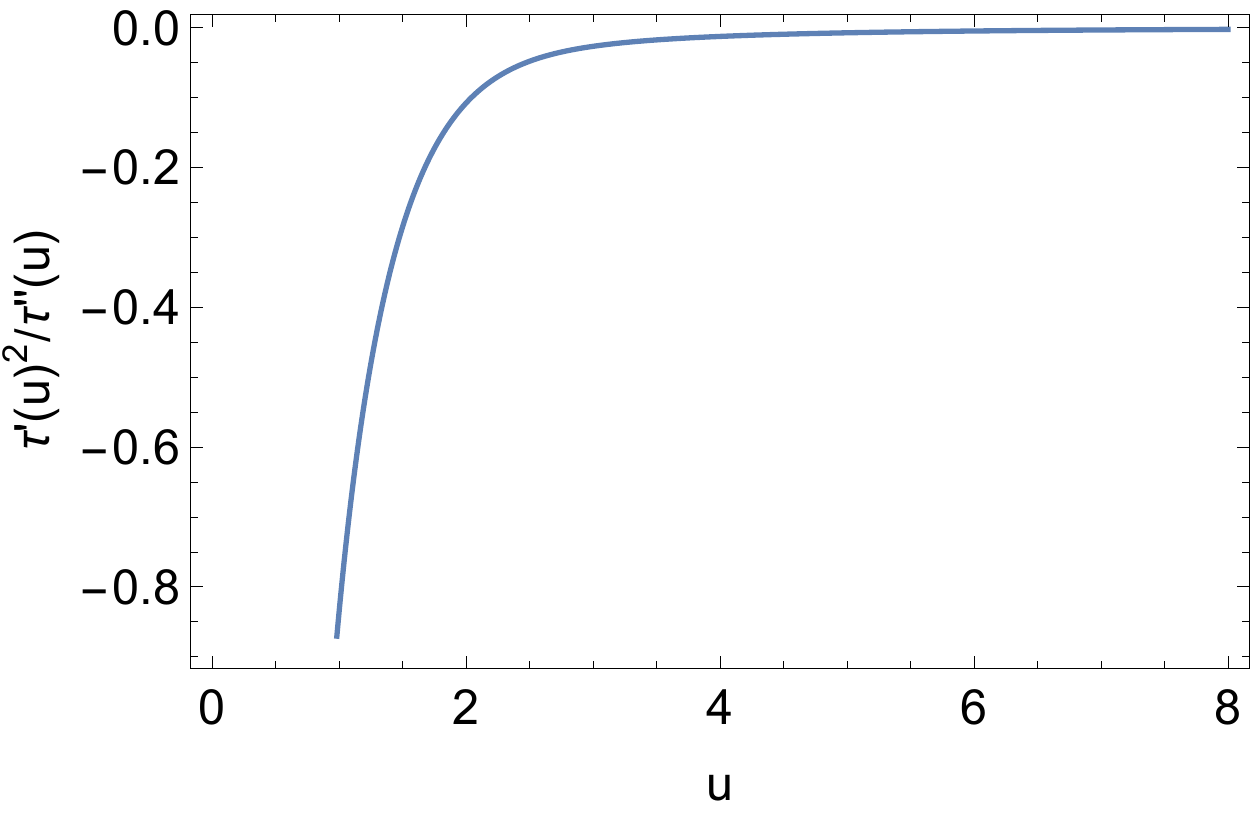}}
\noindent \\
\subfloat[$G(u)$ as a function of time. The dilaton singularity at $r = \infty$ maps to $z = G(u)$ (see text). This implies that the singularity is far from the Poincare horizon $z = \infty$ except at initial time. \label{Fig:Gu}]
{ \includegraphics[width=\linewidth]{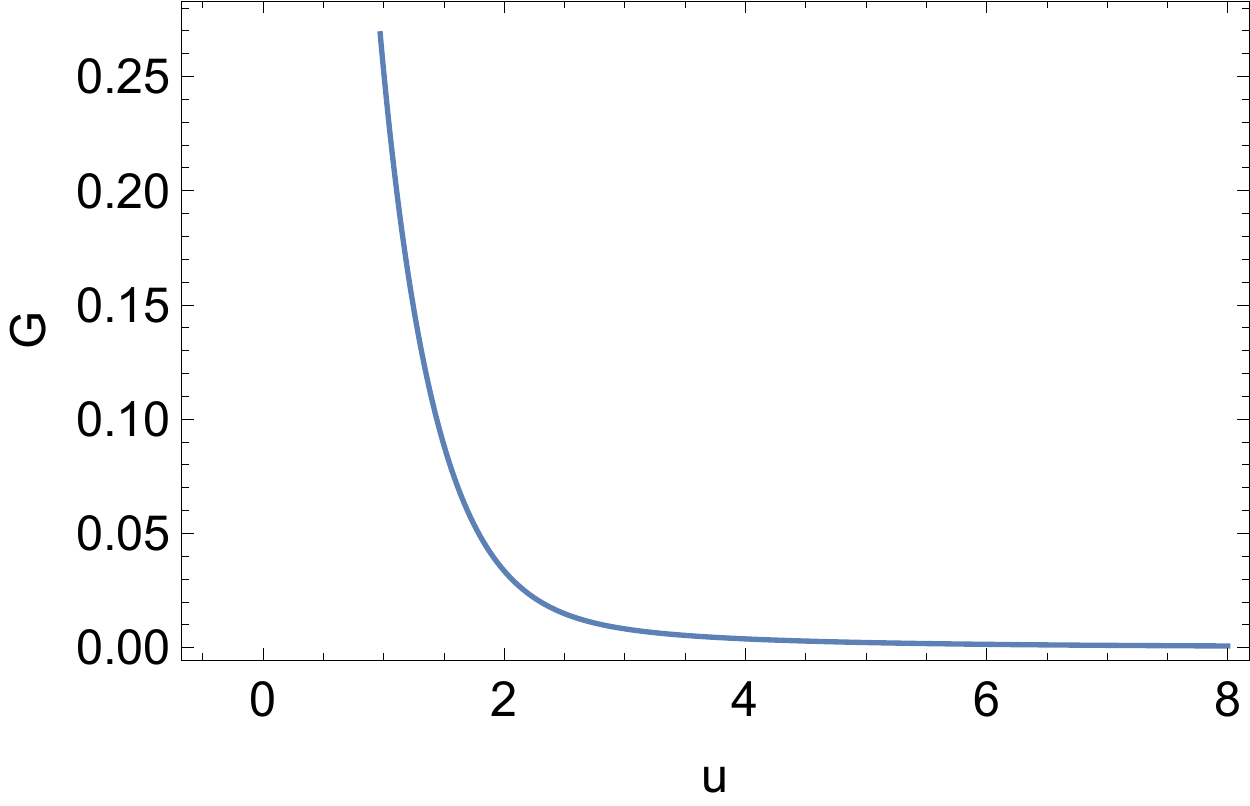}} 

\end{center}
\caption{Here $v_0 = 2.0$ and $\lambda = 0.4$.} 
\end{center}
\end{figure}

\Ayan{Our solution raises an interesting question:  given that the black hole evaporates classically without producing any pathology in the classical gravitational fields, how can we recover the information of the initial conditions from the asymptotic late time behavior. As evident from Fig. \ref{Fig:ChargesPhase1}, all $SL(2,R)$ charges diverge at late time while their Casimir vanishes. We can fit the late time behavior to an exponential proportional to $\exp (a\,u)$ extremely well} \lata{(with an adjusted $R$ square = 0.99)}. \Ayan{It turns out that all $SL(2,R)$ charges (and thus any linear combination of them) grow exponentially with the same exponent $a$. This indeed implies that the Casimir (proportional to $Q^2 - Q^+ Q^-$) vanishes at late time. We conclude that the exponent $a$ is $SL(2,R)$ invariant and an observable.}\footnote{Note that we can measure $\tau(u)$ up to an overall time-independent $SL(2,R)$ transformation simply via a probe coupling to an operator $\tilde{O}$ of the $NAdS_2$ holographic theory. The retarded two-point function of $\tilde{O}$ can be obtained from the linear response by varying the moment of probing in each repeat of the experiment -- this yields  $\tau(u)$ via the conformal map of this propagator to a thermal state with $\beta = 2 \pi$. Irrespective of the initial $SL(2,R)$ frame which can be changed via time-independent $SL(2,R)$ transformation on $\tau(u)$, any linear combination of the late-time $SL(2,R)$ charges will exponentially diverge as $\exp( a\,u)$ with the same invariant exponent $a$.} \Ayan{The initial conditions are labelled by two parameters: (i) the velocity $v_0$ and (ii) the initial mass of the black hole. The total conserved energy determines $v_f$, the terminal velocity of the particle as discussed above. We can then expect that the initial conditions can be recovered completely from the exponent $a$. This is indeed the case as shown in Fig. \ref{Fig:avsvo} where we have plotted how $a$ changes with $v_0$ for a fixed unit initial mass of the black hole. The plot suggests that $a$ grows monotonically with $v_0$. It will be interesting to study the sensitivity of the final state to initial conditions in this context. We leave this for the future.}
\begin{figure}
 \centering
\includegraphics[width=\linewidth]{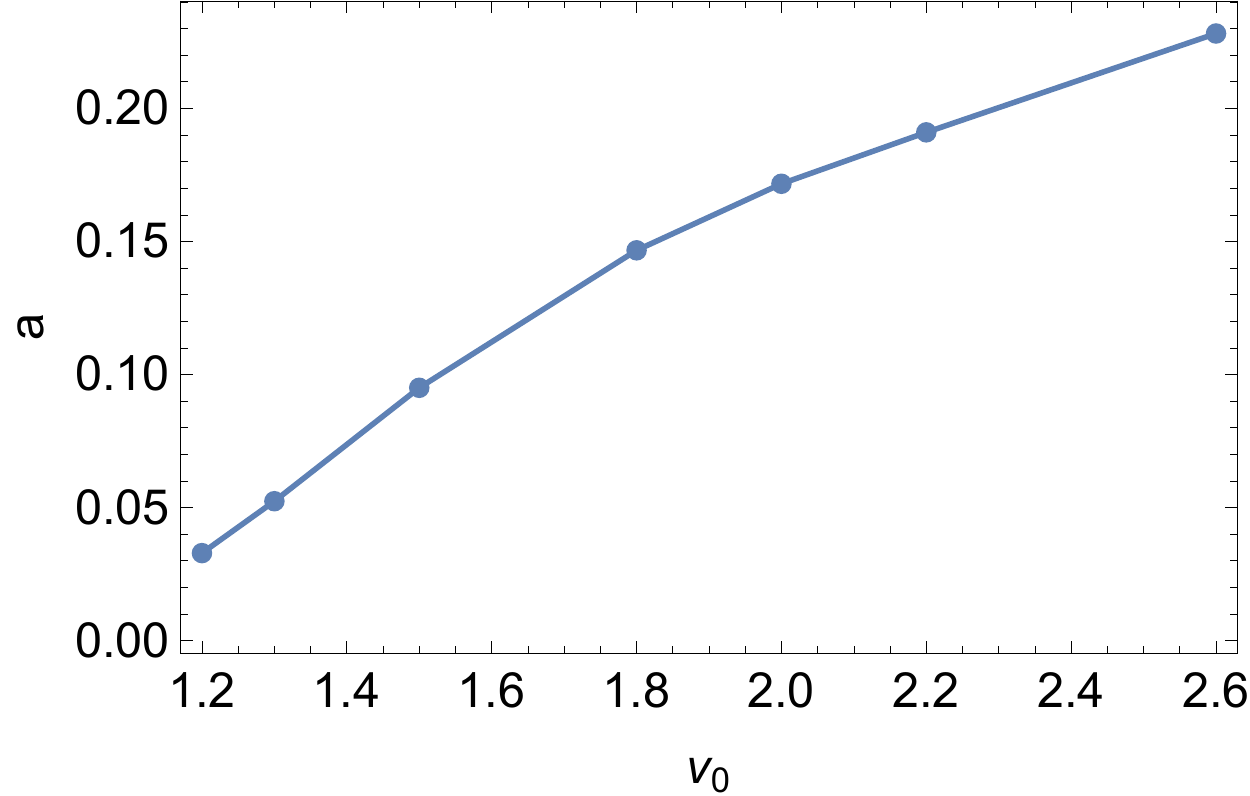}
 \caption{$a$, the exponent for late-time growth of $SL(2,R)$ charges as a function of $v_0$ for $\lambda = 0.4$ and fixed unit initial mass of the black hole. Note that $a$ grows monotonically with $v_0$.}\label{Fig:avsvo}
\end{figure}

\subsubsection{Illustrative examples of phase two behavior}
The second phase appears for $v_0 < v_{c}(\lambda)$. {In this case the mass of the black hole changes sign after finite time before vanishing asymptotically. The total energy is transferred fully to the kinetic energy of the particle or to the confining potential energy depending on whether the total energy is positive or negative, respectively.} We first study the representative case of $v_0 = 0.9$ and $\lambda = 0.4$ {when the total conserved energy is negative}. In this case, the mass of the black hole indeed becomes zero after finite time and then becomes negative (i.e. $H_{sch}$ becomes positive) before vanishing at long time as shown in Fig. \ref{Fig:EnergiesPhase2}. The kinetic energy of the particle goes to zero at late time implying that the particle comes to a full stop after travelling a finite distance, and the total energy gets transferred instead to $H_{int}$, the self-consistent confining potential energy. The gravitational $SL(2,R)$ charges {saturate to constant values in the far future} as shown in Fig. \ref{Fig:ChargesPhase2}. {We observe that although the $SL(2,R)$ charges behave differently from the previously discussed example, $\tau$, $\tau'$, $\tau''$, $\tau'''$ and $G$ behave similarly as functions of $u$.}

\begin{figure}
\begin{center}
\begin{center}
\subfloat[Plot of energies as function of time: Note that the total energy $H_{tot}$ is conserved after initial kick and is finally transferred to $H_{int}$, the confining potential energy. The mass of the black hole $M = - 2 H_{sch}$ becomes negative after finite time and then eventually vanishes at long time. \label{Fig:EnergiesPhase2}]
{ \includegraphics[width=\linewidth]{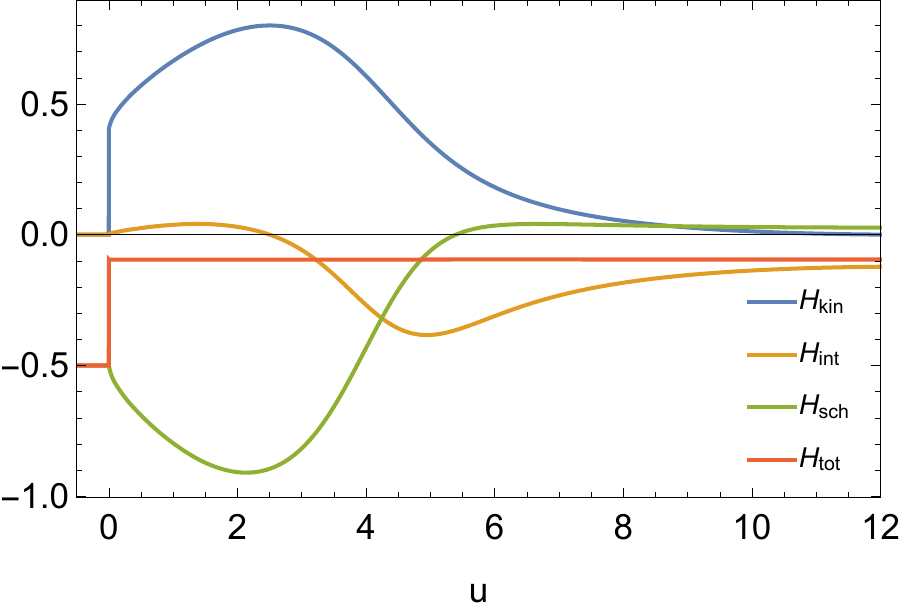}} 
\\
\subfloat[Plot of $SL(2,R)$ charges as a function of time. Note that they saturate to finite values.\label{Fig:ChargesPhase2}]
{  \includegraphics[width=\linewidth]{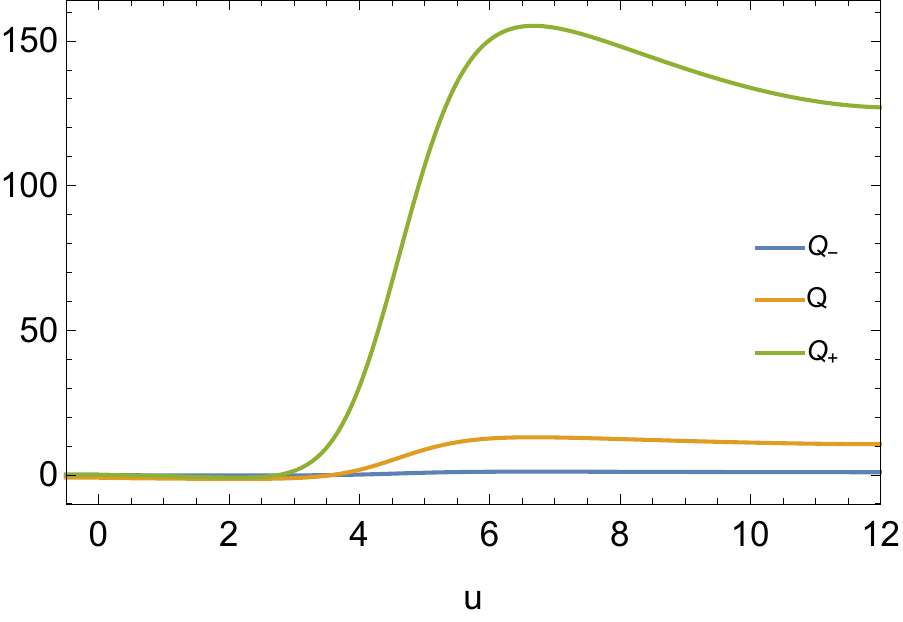}}
\end{center}
\caption{The plots for energies and $SL(2,R)$ charges for $v_0=0.9$ and $\lambda=0.4$.} 
\end{center}
\end{figure}

We readily observe from Fig. \ref{Fig:XuXthPhase2} that $X(u)$ saturates to a finite value at large time implying full stopping. Also $O(u)$ saturates to a finite value at long time (see Fig. \ref{Fig:OuPhase2}) so that indeed $H_{int} = - (\lambda/4)XO$ can also become a constant at long time. On the other hand $O_{th}(\tau(u))$ diverges (see Fig. \ref{Fig:OthPhase2}) but since $\tau'$ decays faster, it is consistent with the product $H_{int} = - (\lambda/4)\tau' X_{th} O_{th}$ going to a constant. Note that $O(u)$ is positive at long time and the final value of confining potential energy is negative as claimed before.

\begin{figure}
\begin{center}
\begin{center}
\subfloat[Plot of $X(u)$ and $X_{th}(\tau(u))$: Both of them saturate to constant values at late time. The particle stops at a finite distance from the origin. \label{Fig:XuXthPhase2}]
{\includegraphics[width=\linewidth]{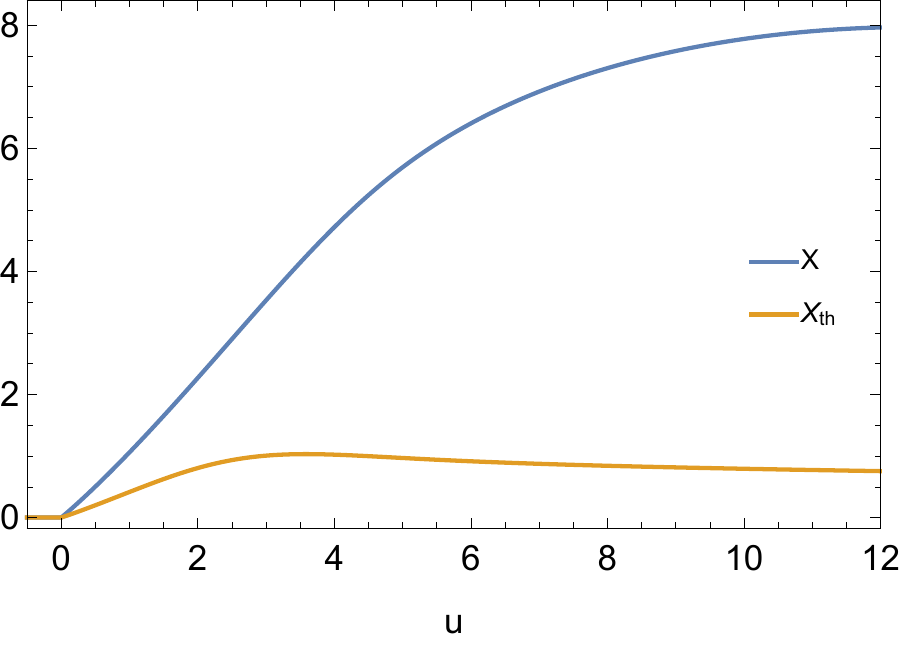}}\\ 
\subfloat[Plot of $O(u)$: $O(u)$ saturates to a constant value at late time so that $H_{int} \propto X(u) O(u)$ also saturates to a constant. \label{Fig:OuPhase2}]
{ \includegraphics[width=\linewidth]{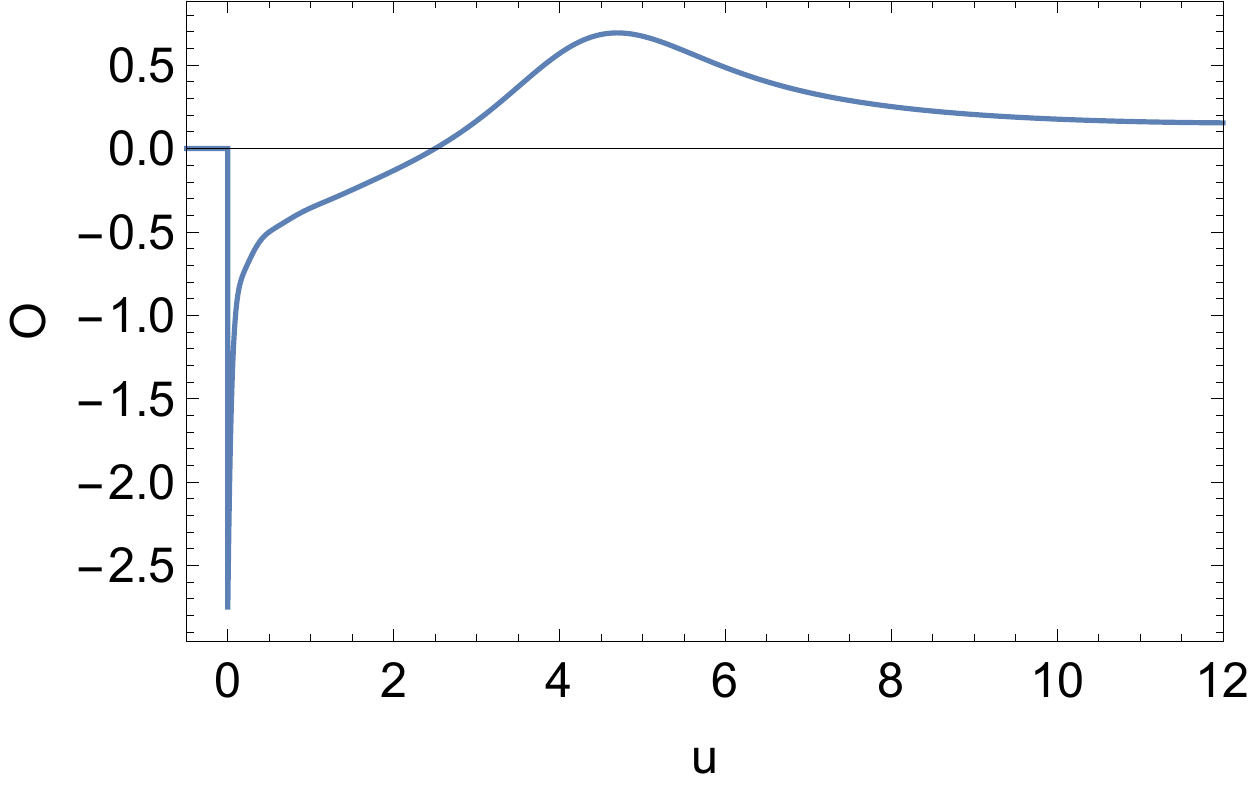}}
\\
\subfloat[Plot of $O_{th}(\tau(u))$: Note $O_{th}(\tau(u))$ diverges at late time since $X_{th}(\tau(u))$ saturates to a constant value. However, $H_{int} \propto \tau'(u) X_{th}(\tau(u)) O_{th}(\tau(u))$ also saturates to a constant in this frame because the decay of $\tau'$ compensates for the growth of  $O_{th}(\tau(u))$. \label{Fig:OthPhase2}]
{ \includegraphics[width=\linewidth]{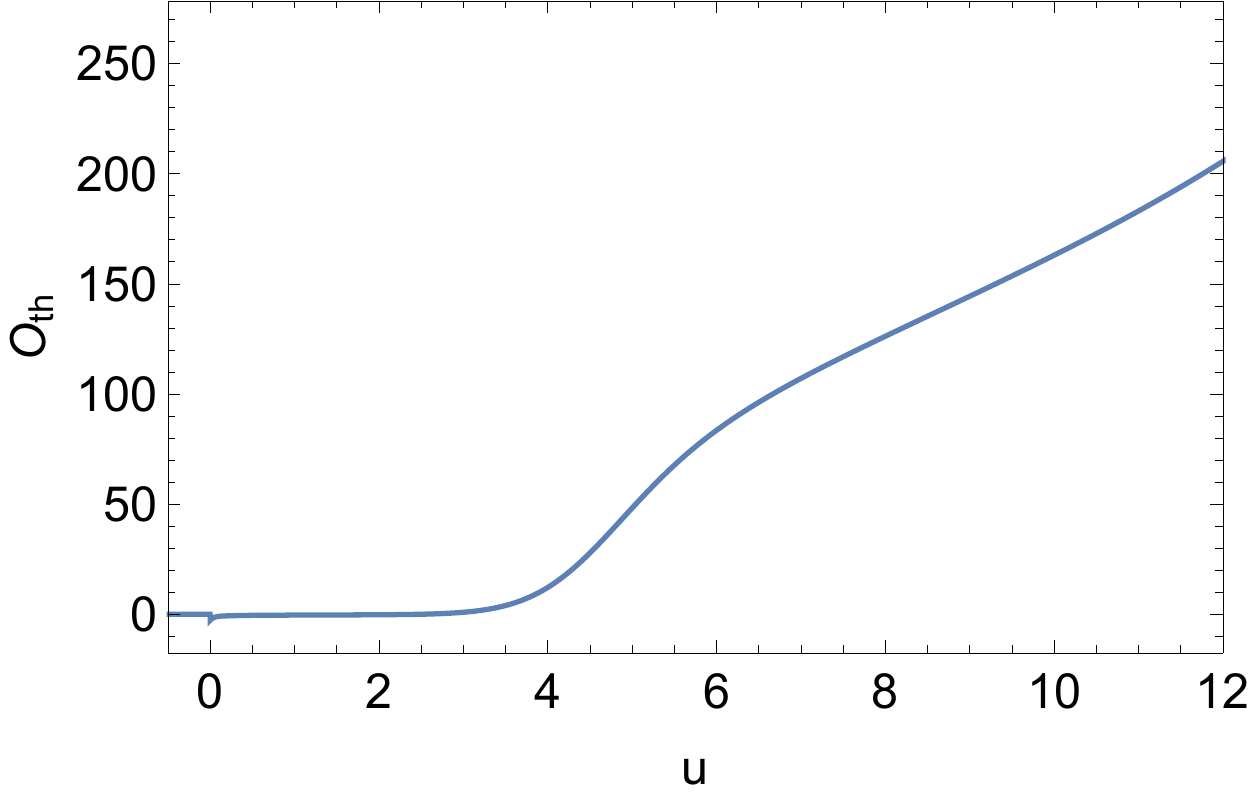}} 
\end{center}
\caption{The plots of sources and responses for $v_0=1.4$ and $\lambda=0.4$.}
\end{center}
\end{figure}

The case of $\sqrt{M_0/m_i} < v_0 < v_c(\lambda)${, when the total conserved energy is positive,} is slightly more complicated. Let us study what happens with $\lambda = 0.5$. The case of $v_0 = 2.0$ corresponds to the first phase and is similar to what has been discussed above. When $v_0 = 1.1$, the final transfer of energy still goes to the kinetic energy of the particle because we choose $\sqrt{M_0/m_i} = 1.0$, but $H_{sch}$ crosses zero \textit{twice} before finally vanishing from below as illustrated in Fig. \ref{Fig:OtherPhases} (see the inset plot). Therefore, for an intermediate time period {the black hole mass is negative}. 

Our results suggest that for $\sqrt{M_i/m_0} < v_0 < v_c(\lambda)$ corresponding to values of $v_0$ for which the total energy should be transferred finally to the particle kinetic energy, the mass $M(u)$ crosses zero an even number of times before finally vanishing from above. The case of $v_0 = 1.1$ just mentioned is illustrated further in Fig. \ref{Fig:v01.1lambda0.5}. For $v_0 < \sqrt{M_i/m_0}$, our results are consistent with odd number of zero crossings of $M(u)$ before its final disappearance. Interestingly, the case of $v_0 = \sqrt{M_i/m_0}$, where $H_{int}$, $H_{kin}$ and $M(u)$ all disappear finally corresponds to a single zero crossing of $M(u)$. However, we warn the reader that since the amplitude of the oscillation of $M(u)$ reduces significantly after each zero crossing of $M(u)$, it is not easy to establish the number of zero crossings definitely numerically as this will require higher precision and also longer time simulations.
The nature of phase transition between the two phases merits a detailed study. The order parameter of this transition is simply the inverse of the crossing time which is the smallest value $u^*$ when $M(u^*) = 0$. Since $M(u)$ never crosses the origin and is positive definite at any finite value of $u$ for $v_0 > v_c(\lambda)$,  the order parameter vanishes in the first phase. In the second phase, the order parameter is finite leading to the vanishing of the black hole mass at $u = u^*$ as discussed above. However, in order to study the phase transition carefully, we need to simulate the full system for very long time for $v_0$ close to $v_c$ which is a significant numerical challenge as mentioned above. We leave this for the future. 

{In all cases we have studied, the dilaton is well-behaved in the entire physical patch covered by the $r$ and $u$ coordinates.}
\\
\begin{figure}
 \centering
\includegraphics[width=\linewidth]{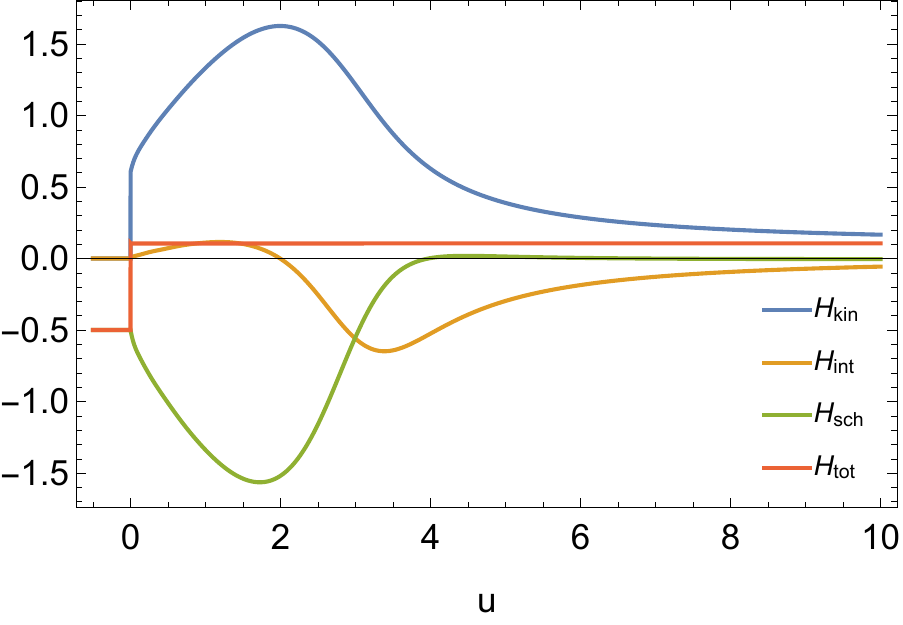}
 \caption{The four energies in the case of $v_0 = 1.1$ and $\lambda = 0.5$. The double crossing of $H_{sch}$ about zero is hard to discern here, so one can refer to the inset plot in Fig. \ref{Fig:OtherPhases}. The final transfer of energy goes to the kinetic energy of the particle.}\label{Fig:v01.1lambda0.5}
\end{figure}
\begin{figure}
 \centering
\includegraphics[width=\linewidth]{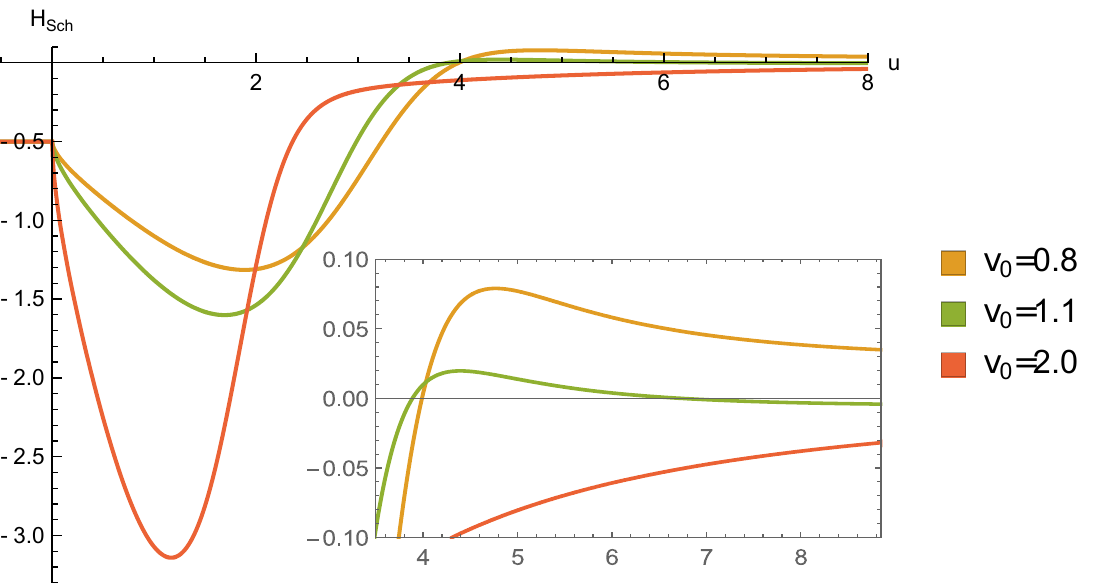}
 \caption{Different phases for $\lambda = 0.5$ as can be seen from the behavior of $M(u) = - 2 H_{sch}(u)$ for various choices of initial velocities. The inset plot shows that multiple crossings of zero is possible for $M(u)$ when the total conserved energy is positive.}\label{Fig:OtherPhases}
\end{figure}
\subsection{Remarks on the second law}
The eventual vanishing of the black hole mass which is extracted completely in the form of the particle's kinetic or potential energy naturally begets the question of consistency with the second law of thermodynamics.  In fact, we have discussed in Section \ref{dilpf} that JT gravity possesses a second law formally and in case of the holographic quench we have explicitly verified the monotonic growth of entropy interpolating between the thermal limits at early and late times, which implies that the final black hole mass should be greater than the initial value. 

In order to track the second law, we first need to find an appropriate smooth null curve (geodesic). In the case of the holographic quench, this curve was the event horizon, which interpolates between initial and final thermal horizons. In the present case, it is not possible to define such an event horizon within the physical patch because the system never thermalizes (so a future boundary condition is meaningless) or the mass of the black hole becomes negative. Instead we can define a dynamical horizon which can be readily stated in the $\rho$ and $\tau$ coordinates of the black hole with fixed mass $M = 1$, which maps to the physical $r$ and $u$ coordinates via \eqref{EFdiffeo}. This choice is simply the horizon of the $M=1$ black hole given by $\rho = 1$ which maps to the horizon of the physical varying mass black hole solution given by \eqref{EFMu} via
\begin{equation}\label{Eq:physhorizon}
r = \frac{1}{\tau'(u) \left(1 + \frac{\tau''(u)}{\tau'^2(u)}\right)}.
\end{equation}
This dynamical horizon is causal since $\tau(u)$ is causal as well. Also it coincides with the initial thermal horizon at $r = 1$. Therefore, following our earlier discussion, the initial value of the dilaton on the horizon can reproduce the initial thermal entropy.


Remarkably, as shown in Fig. \ref{Fig:RunawayHorizon}, we find that the dynamical horizon $r(u)$ runs away to infinity almost at the same time as the black hole mass stops increasing monotonically as can be observed by comparing with  Fig. \ref{Fig:OtherPhases}. Clearly, it is not possible to formulate the second law beyond this limiting value of $u$. This is irrespective of whether the system is in the first or the second phase, i.e. whether the black hole mass vanishes with or without crossing zero after the horizon runs away to $r = \infty$.\footnote{It turns out that in case of the holographic quench, the dynamical horizon \eqref{Eq:physhorizon} runs away to infinity in finite time as well. However, we can utilize the event horizon that cannot be meaningfully defined in the semi-holographic case. This runaway behavior in the holographic case can be associated to the formation of $SL(2,R)$ hair, which is the relative $SL(2,R)$ frame rotation between the initial and final thermal states (it is an observable as mentioned before). This is similar to the semi-holographic case where the runaway is associated to subsequent approach to a quantum attractor that can be characterized by an $SL(2,R)$ invariant exponent as discussed below.}

We note that the $r$ and $u$ coordinates are actually a double cover of the physical patch simply because $r\rightarrow -r$ keeps the metric \eqref{EFMu} invariant. This other sheet, where $r$ is negative, is actually not physical -- the dilaton is complex here. As shown in Fig. \ref{Fig:RunawayHorizon}, after the horizon runs away to $r = \infty$, it re-emerges in this non-physical patch. It is also important to note that since the states of the theory belong to the coset space $Diff/SL(2,R)$, it is important to choose bulk coordinates which coincide with the physical time $u$ at the boundary. Therefore the horizon should be studied in the ingoing Eddington-Finkelstein $r$ and $u$ coordinates (and not the $\rho$ and $\tau$ coordinates of the fixed unit mass black hole).
\begin{figure}
 \centering
\includegraphics[width=\linewidth]{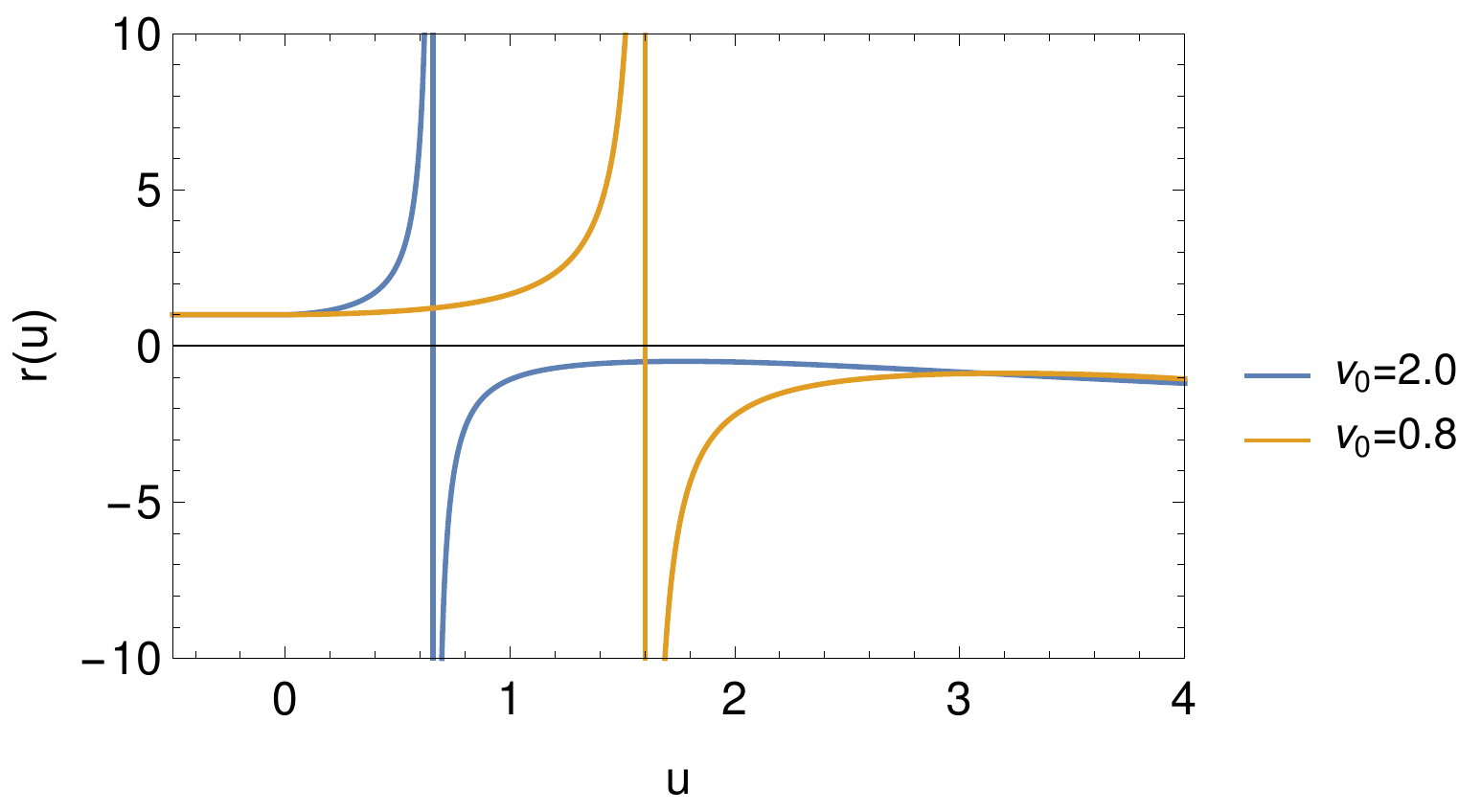}
 \caption{The behavior of the horizon $r(u)$ given by \eqref{Eq:physhorizon} for $\lambda = 0.5$, and different initial velocities of the particle, namely $2.0$ and $0.8$. In the former, the system is in the first phase and in the latter, it is in the other phase. Note that the initial horizons coincide because we start with the same black hole mass. The horizon runs away to infinity irrespective of whether the black hole mass becomes negative or not. Also it happens roughly when the black hole mass stops increasing monotonically as evident from comparison with Fig. \ref{Fig:OtherPhases}. After this the horizon emerges on the other (unphysical) sheet (see text).}\label{Fig:RunawayHorizon}
\end{figure}

It has been shown in \cite{Kurkela:2018dku} that in semi-holographic setups there exists a second law for the full system provided there exists entropy currents in each individual (perturbative and holographic) subsystems. In the special case where only one highly energetic degree of freedom is coupled to a large holographic system as in \cite{Ecker:2018ucc}, the holographic system alone captures the second law as expected from equipartition of energy. This is true for higher dimensional holographic setups as demonstrated in \cite{Ecker:2018ucc} but not in the present case. Although we are unable to provide a quantitative argument as of yet, we can readily see that the origin of this discrepancy lies in entanglement between the two sectors, which if taken into account properly should be able to rescue the second law. 

As discussed in Section \ref{sec:phase1}, the initial conditions of the particle can be recovered from the $SL(2,R)$ invariant exponent $a$ governing the growth of the gravitational bulk $SL(2,R)$ charges and the total conserved energy.\footnote{Of course this is in the case when the system is in the first phase. In the other case when the $SL(2,R)$ charges stabilize in the far future, the knowledge of the initial velocity of the particle can be retrieved from the relative rotation between the initial and final $SL(2,R)$ frames which is an observable.} Therefore the information of one degree of freedom is not lost, but rather encoded in the macroscopic behavior of the much larger holographic system. This indicates a form of macroscopic entanglement which allows information in a tiny system (of measure zero) to be visible in terms of the macroscopic coarse-grained dynamics of the larger system.  It is well known from quantum information theory that entanglement contributes to the second law and can account for the apparently bizarre reverse flow of heat from a cold to a hot body which have large initial mutual entanglement \cite{PhysRevE.77.021110}. The growth of entanglement can account for the second law most likely in our case. Quantitatively this could be perhaps understood from the solution in the unphysical sheet (where the horizon is at a negative value of $r$). We will leave a more precise formulation for the future.

Finally, our model is somewhat novel in the context of semi-holography because the total entropy is not just the sum of the two individual entropies in this case (as expected from classical statistical mechanics, see \cite{Kurkela:2018dku} for details) but also involves the crucial contribution from entanglement between the two subsystems especially in the limit when the holographic subsystem has larger number of degrees of freedom.

\section{Conclusions}
In this work we have developed a concrete algorithm for constructing solutions of JT gravity coupled to non-conformal matter and have constructed explicit bulk solutions corresponding to time-dependent irrelevant deformations in the dual theory. We find that such perturbations act like pumps increasing the mass of the black hole as expected.

We also construct a non-geometric semi-holographic string model for trapped strongly interacting impurities. The $NAdS_2$ holographic theory depicts the mutual strong interactions of the localized impurities. It couples to the position of one displaced impurity which thus acts as a self-consistent source of an irrelevant operator dual to a bulk field. This operator in turn gives rise to a confining force on the impurity. The model has a total conserved energy. The impurity gets displaced  from the confining center due to a kick from the thermal medium. We have studied how it moves in response to such an impulse.  

Keeping the mutual coupling $\lambda$ fixed, we find two distinct phases. In the first phase occurring for higher initial velocity $v_0 > v_c(\lambda)$, the impurity extracts all energy from the bulk and finally the total energy gets transferred to its kinetic energy with a terminal velocity smaller than the initial velocity. Furthermore, the mass of the bulk black hole remains always positive vanishing from above. Although the black hole mass vanishes, the solution does not reach vacuum. The $SL(2,R)$ charges grow exponentially (although the Casimir tends to vanish) and the dilaton has non-trivial time-dependence. If $v_0 < v_c(\lambda)$, the total energy is transferred either to the particle kinetic energy if the total conserved energy is positive, or otherwise to the self-consistent confining potential energy in which case the particle comes to a full stop.  

In reality, the impulse kicking the impurity initially from the center should originate from the thermal medium so $v_0$ should be on average of the order of $\sqrt{ 2k_B T/m_{i}}$ where $m_i$ is the mass of the impurity. If  $\sqrt{2 k_B T/m_{i}} > v_{c}(\lambda)$, then the dynamics is as in the first phase. Therefore, impurities can travel long distances while remaining correlated with each other and attaining terminal velocities less than $\sqrt{ 2k_B T/m_{i}}$ at long time.

\Ayan{The key feature of $NAdS_2$ semi-holographic systems (with a total conserved energy) that we find here is that the black hole can lose its mass to the dynamical source at the boundary or to the mutual self-consistent potential energy over a large period of time. Furthermore, the solutions are non-pathological. As opposed to higher dimensional setups, here the information about the impurity can be recovered from the macroscopic parameter determining the long-term attractor behavior of the larger holographic system, namely the $SL(2,R)$ invariant exponent governing the growth of the bulk $SL(2,R)$ charges when  $v_0 > v_c(\lambda)$. This implies that in order to understand the second law we need to take into account entanglement between the impurity and coarse-grained degrees of freedom of the holographic system.} 

\Ayan{{Considering a chain or a lattice of such $NAdS_2$ holographic systems, we can answer many interesting questions at the interface of quantum information and many-body dynamics. In the future, it will be interesting to probe existence of possible bounds on the response of the final state to changes in initial conditions, investigate possibility of chaotic behavior far away from equilibrium and see if we can utilize such systems for quantum tasks such as quantum error correction.}}

\begin{acknowledgments}
We thank Saumen Datta, Daniel Grumiller,  Arnab Kundu, R. Loganayagam, Gautam Mandal, Shiraz Minwalla,   Giuseppe Policastro, Anton Rebhan and Sandip Trivedi for very helpful comments and discussions. We thank Daniel Grumiller,   Giuseppe Policastro, Anton Rebhan and Alexandre Serantes for comments on the manuscript. \Ayan{We thank Tanay Kibe and Hareram Swain for providing the first version of Fig 9.} A.\ Mukhopadhyay acknowledges support from the Ramanujan Fellowship and Early Career Research Award of SERB of DST, India and also the new faculty initiation grant of IIT Madras. A.\ Soloviev is supported by the Austrian Science Fund (FWF) doctoral program W1252.  The  work  of  L.K. Joshi  has been  supported  in  part by the Infosys foundation for the study of quantum structure of spacetime.
\end{acknowledgments}

\bibliography{NAdS2}

\end{document}